\documentclass[journal=jpcbfk,manuscript=article]{achemso}

\usepackage[utf8]{inputenc}
\usepackage[T1]{fontenc}
\usepackage{amsmath}
\usepackage{xr-hyper}
\usepackage{soul}
\usepackage[colorlinks,allcolors=blue]{hyperref} 
\usepackage{braket}
\usepackage{numprint}
\usepackage{filecontents}
\usepackage[dvipsnames]{xcolor}

\makeatletter
\newcommand*{\addFileDependency}[1]{
\typeout{(#1)}
\@addtofilelist{#1}

\IfFileExists{#1}{}{\typeout{No file #1.}}
}\makeatother

\newcommand*{\myexternaldocument}[1]{
\externaldocument{#1}
\addFileDependency{#1.tex}
\addFileDependency{#1.aux}
}

\myexternaldocument{suppinfo}
\SectionNumbersOn

\author{Alessia Valzelli}
\affiliation[]{Dipartimento di Ingegneria dell’Informazione, Università degli Studi di Firenze, 50139 Firenze, Italy}
\email{alessia.valzelli@unifi.it}
\alsoaffiliation[]{Dipartimento di Fisica e Astronomia, Università degli Studi di Firenze e CSDC, 50019 Sesto Fiorentino, Italy}
\alsoaffiliation[]{Istituto Nazionale di Fisica Nucleare, Sezione di Firenze, 50019 Sesto Fiorentino, Italy}

\author{Alice Boschetti}
\affiliation[]{European Laboratory for Non-Linear Spectroscopy (LENS), Università degli Studi di Firenze, 50019 Sesto Fiorentino, Italy}
\alsoaffiliation[]{Istituto Nazionale di Ricerca Metrologica (INRiM), 10135 Torino, Italy}

\author{Francesco Mattiotti}
\affiliation[]{University of Strasbourg and CNRS, CESQ and ISIS (UMR 7006), aQCess, 67000 Strasbourg, France}

\author{Armin Kargol}
\affiliation[]{Loyola University New Orleans, Dept. of Physics, New Orleans, LA 70118, USA}

\author{Coleman Green}
\affiliation[]{Loyola University New Orleans, Dept. of Physics, New Orleans, LA 70118, USA}

\author{Fausto Borgonovi}
\affiliation[]{Dipartimento di Matematica e Fisica and Interdisciplinary Laboratories for Advanced Materials Physics,
Università Cattolica, 25133 Brescia, Italy}
\alsoaffiliation[]{Istituto Nazionale di Fisica Nucleare, Sezione di Milano, 20133 Milano, Italy}

\author{G. Luca Celardo}
\affiliation[]{Dipartimento di Fisica e Astronomia, Università degli Studi di Firenze e CSDC, 50019 Sesto Fiorentino, Italy}
\alsoaffiliation[]{Istituto Nazionale di Fisica Nucleare, Sezione di Firenze, 50019 Sesto Fiorentino, Italy}
\alsoaffiliation[]{European Laboratory for Non-Linear Spectroscopy (LENS), Università degli Studi di Firenze, 50019 Sesto Fiorentino, Italy}

\title[Large scale simulations]{Large scale simulations of photosynthetic antenna systems: interplay of cooperativity and disorder}

\abbreviations{GSB,PB,Bchl,2LS,RC,TDM,FMO,LHI,LHII,NHH,HH,DH}

\begin{document}
\renewcommand{\i}{{\mathrm{i}}}

\begin{tocentry}

    \centering
    \includegraphics{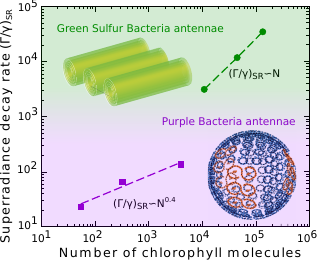}

    \label{For Table of Contents Only}

\end{tocentry}

\begin{abstract}
Large scale simulations of light-matter interaction in natural photosynthetic antenna complexes containing more than one hundred thousands chlorophyll molecules, comparable with natural size, have been performed. Photosynthetic antenna complexes present in Green sulfur bacteria and Purple bacteria have been analyzed using a radiative non-Hermitian Hamiltonian, well known in the field of quantum optics, instead of the widely used dipole-dipole Frenkel Hamiltonian. This approach allows  to study ensembles of emitters beyond the small volume limit (system size much smaller than the absorbed wavelength), where  the Frenkel Hamiltonian fails.   When analyzed on a large scale, such structures display  superradiant states much brighter then their single components. An analysis of the robustness to static disorder and dynamical (thermal) noise, shows that  exciton coherence in the whole photosynthetic complex is larger than the coherence found in its parts. This provides evidence that the photosynthetic complex as a whole has a predominant role in sustaining coherences in the system even at room temperature. 
Our results allow a better  understanding of natural photosynthetic antennae and could drive experiments to verify how the response to the electromagnetic radiation depends on the size of the photosynthetic antenna.   
\end{abstract}

\section{Introduction}
\label{sec:intro}

Photosynthesis is a fundamental process able to  capture  Sun's energy and convert it into biochemical energy used to drive cellular processes~\cite{blankenship2021molecular}. Here  we investigate antenna complexes of photosynthetic anaerobic bacteria: the \textit{Chlorobium Tepidum} Green sulfur bacteria (GSB) and  \textit{Rhodobacter sphaeroides} Purple bacteria (PB) which are   among the most efficient photosynthetic complexes in Nature~\cite{damjanovic2000excitation}. Indeed anaerobic bacteria have the ability to harvest sunlight in deep murky waters, where incident light levels are much reduced beyond the already dilute level on land~\cite{strumpfer2012quantum,csener2007atomic}. For instance, Purple bacteria have the ability to exploit extremely weak light sources~\cite{strumpfer2012quantum,csener2007atomic,sawaya2015fast, huh2014atomistic} (less than 10 photons per molecule per second) and some species of Green sulfur bacteria even perform photosynthesis with geothermal radiation from deep-sea hydrothermal vents at about 400$^\circ$C~\cite{greenPNAS}.

Photosynthetic antenna complexes of anaerobic bacteria~\cite{strumpfer2012quantum,csener2007atomic,schulten1:1,schulten1:2,photo,photoT,grad1988radiative,mukamel,srfmo,srrc} are comprised of a network of Bacteriochlorophyll (Bchl) molecules which are typically modelled as two-level systems (2LS) capable of absorbing radiation and transfering the resulting electronic excitation to the reaction center (RC) where charge separation occurs, a process which precedes and drives all other photosynthetic steps. To each 2LS a transition dipole moment (TDM) is associated which determines its coupling with both the electromagnetic field and  with other  chlorophyll molecules. 
Owing to the low solar photon density, photosynthetic aggregates operate in the single-excitation regime, meaning that at most one excitation is present in the system at any  time.
Antenna complexes of photosynthetic bacteria have an internal efficiency of almost $100 \%$ (i.e. nearly each photon absorbed produces a charge separation event in the RC)~\cite{strumpfer2012quantum,huh2014atomistic}. 

A possible origin of this incredible ability of bacterial photosynthetic systems to utilize weak sources of incoherent light and funnel the collected energy to specific molecular aggregates could be brought back to the high level of symmetry and hierarchical organization characterizing the antenna complexes of bacterial photosynthetic organisms~\cite{kassal,kassal2}. Photosynthesis in GSB involves chlorophyll pigments tightly packed in light-harvesting systems with cylindrical shapes, known as chlorosomes~\cite{blankenship2021molecular} and shown in  Fig.~(\ref{modelsGSB}). They are able to absorb the sunlight and transfer the  electronic excitation to other fundamental units, such as the baseplate and the Fenna-Matthews-Olson (FMO) trimer
complex, and finally to the reaction centers (RCs), where the incoming energy is converted into a charged-separated state~\cite{strumpfer2012quantum, huh2014atomistic}. Purple bacteria offer an alternative organization of the antenna system. The main
photosynthetic units of Purple bacteria are the  chromatophores ($ \approx 60$ nm in size)~\cite{csener2007atomic} which contain about $5000$ BChl molecules~\cite{qian2013three,koolhaas1998identification}, and are composed of different antenna complexes: LHI, LHII and the RC~\cite{csener2007atomic,saer2017light,schulten3}. LHI and LHII complexes show a very well ordered disposition of Bchl molecules which are arranged in ring and S-shaped structures.  The energy collected by the chromatophore reaches the RC complex mainly through the
LHI complex which surrounds it (see Fig.~(\ref{modelsPB})). 

The basic components of the photosynthetic antenna complexes of anaerobic bacteria have been widely studied 
both theoretically and experimentally in Refs~\cite{saer2017light,eltsova2016effect,chew2007bacteriochlorophyllide,linnanto2013exciton,malina2021superradiance,molina2016superradiance,ref5}. 
Due to the symmetric arrangement of Bchl molecules these structures  display bright (superradiant) and dark (subradiant) states in their single-excitation manifold~\cite{monshouwer1997superradiance,strumpfer2012quantum}. Bright states are characterized by a giant transition dipole moment (much larger than the single molecule dipole moment), while dark states exhibit a significantly smaller transition dipole moment compared to that of a single molecule. 
In GSB bright states close to the lower excitonic state have been found in cylindrical Bchl aggregates~\cite{gulli2019macroscopic,molina2016superradiance,knoester2006modeling,didraga2004excitons,eisele2009uniform,vlaming2011disorder}. Additionally, in the PB antenna complex it is well-known that both LHI and LHII complexes display bright states close the lower excitonic state~\cite{strumpfer2012excited,csener2007atomic,kassal,kassal2,monshouwer1997superradiance}.

Nevertheless, the cooperative response to light of the entire photosynthetic antenna complexes has not been theoretically studied thus far. Indeed, whether the arrangement of the Bchl molecules in the entire GSB and PB antenna complexes is capable of supporting collective states brighter than the single sub-units in the complex,  is not a trivial question to address. On one side, it is not guaranteed that the symmetry of the entire antenna is capable of supporting a   cooperative response larger than its sub-units. Moreover, the most widely used theoretical model, the Frenkel Hamiltonian~\cite{may2023charge} based on dipole-dipole interaction, becomes ineffective beyond the small volume limit, i.e.~when the system size is comparable with the    wavelength of the absorbed light. For this reason here we employ a radiative Hamiltonian model, well-established  in quantum optics for several decades~\cite{grad1988radiative,gulli2019macroscopic}. The radiative Hamiltonian allows us to explore light-matter interaction of photosynthetic antennae beyond the small volume limit,  where the Frenkel Hamiltonian cannot be used. The  radiative Hamiltonian contains non-Hermitian terms that account for photon losses due to spontaneous emission and accurately describes the effective interaction between molecules. 
Even if the non-Hermitian part is usually considered in the frame of perturbation theory, this approximation breaks down in large systems when resonances overlap\cite{zelevinsky}.

In this study, we thus conduct a large-scale analysis of Green Sulfur Bacteria and Purple Bacteria photosynthetic complexes. Specifically, we examine the GSB chlorosome, which comprises three adjacent concentric cylinders containing more than 10$^5$ chlorophyll molecules, see Fig.~(\ref{modelsGSB}), and the entire chromatophore present in Purple Bacteria, containing approximately \numprint{5000} chlorophyll molecules, see Fig.~(\ref{modelsPB}). 

We find that cooperative effects in the whole photosynthetic complex are enhanced with respect to its smaller sub-units even when realistic levels of disorder and thermal noise are considered.

\section{Models and Methods}
\label{models-methods}

The positions and the orientations of the Bchl molecules in both GSB and PB, exhibit a high degree of hierarchical order and symmetry. However, the connection between these structures and their functionality remains an open question. Recent research by some of the authors of this manuscript has demonstrated that the natural arrangement of Bchl molecules in GSB cylindrical structures ensures the presence of cooperative effects  even at room temperature~\cite{gulli2019macroscopic}. In this study, we analyze much larger complexes containing  an order of magnitude more Bchl molecules
and thus more close to natural sizes. 

Below we describe the specific complexes employed in this manuscript and the different Hamiltonian models used to analyze their response to the electromagnetic field. We also describe  the theoretical methods used to analyze the robustness of their optical properties to  static disorder and  thermal noise.

\subsection{Antenna complexes in Green Sulfur Bacteria (GSB)}
\label{subsec:GSBmodel}
The most common pigments found in the
GSB antenna~\cite{orf2013chlorosome,huh2014atomistic} are Bchls~\textit{c}, \textit{d}, or \textit{e}, along with carotenoids, and to a lesser extent, Bchls~\textit{a}~\cite{gunther2016structure,linnanto2013exciton,linnanto2008investigation}, although  Bchls~\textit{c} is the most abundant pigment~\cite{pedersen2010model,frigaard2006chlorosomes}. 
Pigment organization and orientations of the GSB chlorosome have been studied by using various spectroscopic methods, such as infrared and resonance Raman studies, solid state NMR and cryo-EM, that have revealed that it can be formed by pigments assembling in rod-like (cylindrical) aggregates with lateral lamellae~\cite{linnanto2013exciton,linnanto2008investigation,oostergetel2007long,pvsenvcik2004lamellar,pvsenvcik2006internal,psencik2009structure,gunther2016structure}.
 The specific structure  are strongly dependent on the growing  and the environmental conditions~\cite{gulli2019macroscopic, hohmann2005ultrastructure, oostergetel2010chlorosome}. In particular,  \textit{Chlorobium Tepidum} triple mutant antenna complexes present cylindrical structures that can contain between \numprint{50000} to \numprint{250000} Bchl\textit{c}. These structures typically range in size from $100$ to $200$ \mbox{nm} in length  with widths and depths varying between $60$ and $100$ nm~\cite{orf2013chlorosome,staehelin1978visualization,staehelin1980supramolecular,cohen1964fine}.  
 
Although natural antennae of GSB exhibit considerable variation in size and in the types of Bchl aggregates present (such as lamellae and cylindrical structures), here
 we model the antenna  using only  cylindrical structures  to capture the primary natural features. 
 The cylindrical structures in GSB are made of concentric single wall cylinders, which are observed to lie adjacently to each other on the baseplate. 
 
\begin{figure}[!ht]
    \centering
    \includegraphics[width=\columnwidth]{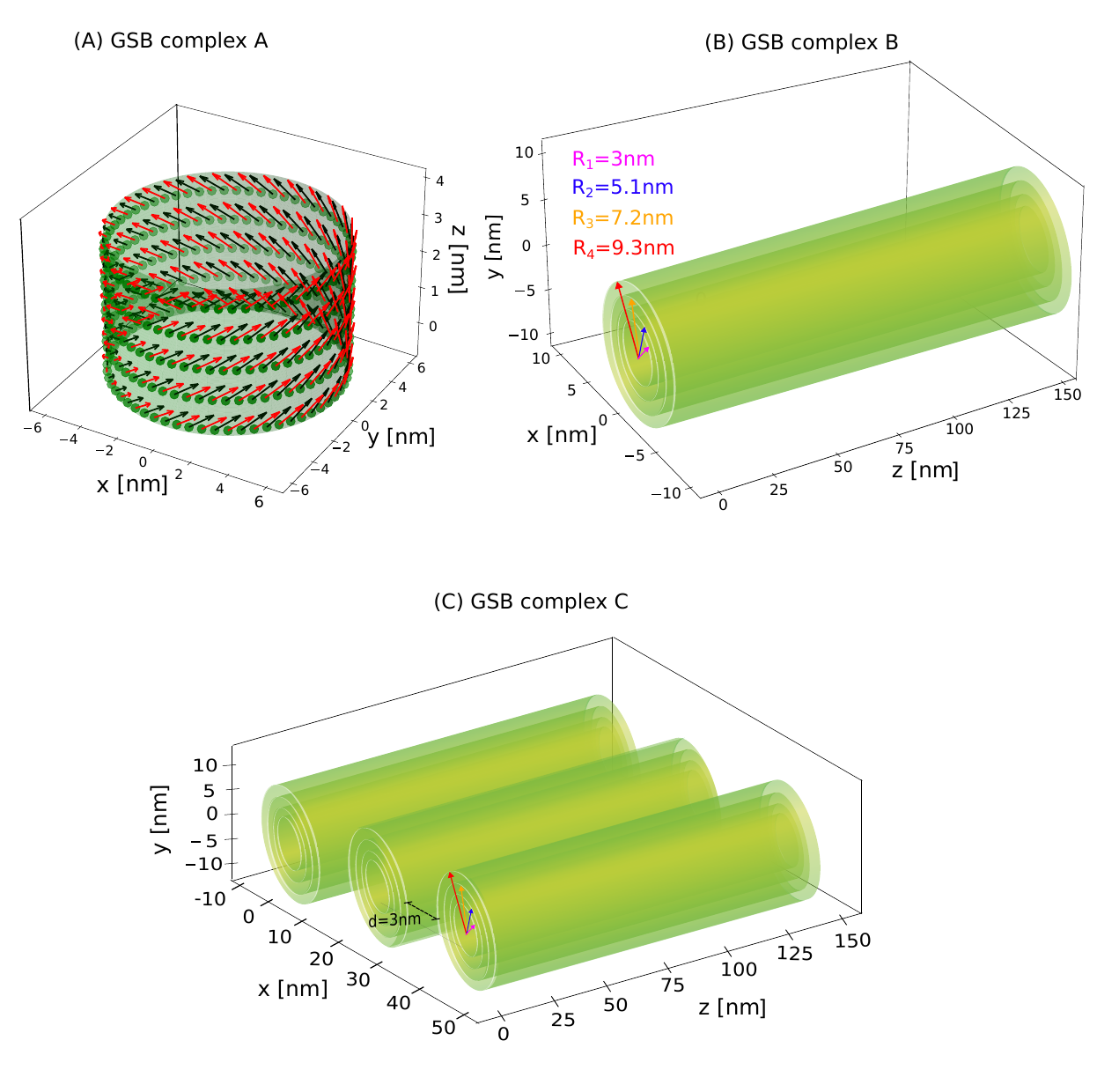}
    \caption{{\it Architecture of GSB light-harvesting complexes.}
    Cylindrical complexes representing  the Green sulfur bacterium bchQRU triple mutant antenna system. (A) Complex A: a section of a single cylinder with a radius $R = 6 \ \mbox{nm}$ and 5 rings is represented. Each Bchl molecule is associated to a  transition dipole moment represented by an arrow with a well defined position and orientation.   TDM belonging to the same ring are tilted
by small alternating angles $\pm \alpha$ out of the lattice plane as a result of the syn-anti stacking~\cite{gunther2016structure,ref1,ref4}. This alternation is so small that optically it may be neglected~\cite{gunther2016structure}. In particular 
    black (red)  arrows on the same ring characterizes dipoles that point inward (outward) with respect to the tangent plane of the cylinder by a small angle $\alpha=\pm 4^{\circ}$, for more details see Ref.s~\cite{gulli2019macroscopic,gunther2016structure}. (B) Complex B:  representation of the structure with four concentric cylinders with  radii of $3-5.1-7.2-9.3 \ \mbox{nm}$ and containing respectively $30-51-72-93$ dipoles per ring. 
On each wall the positions and the orientations of the transition dipole moments are the same as in panel (A). (C) Complex C: model formed by three adjacent cylindrical aggregates with four concentric rolls each. }
    \label{modelsGSB}
\end{figure}

The model that we consider is shown in Fig.~(\ref{modelsGSB}). It is inspired by   the antenna complexes found in the Green sulfur bacterium \textit{C. Tepidum} bchQRU triple mutant type, whose chlorosomes exhibit a much more regular geometry with respect to the wild type. Nevertheless, both types exhibit very similar optical properties and demonstrate similar cooperative effects~\cite{gulli2019macroscopic,  ganapathy2009alternating}. 
Below, we describe, in order of hierarchical complexity, the geometry of three structures utilized in this study to model the cooperative effects occurring in the antenna systems of photosynthetic GSB.

\textbf{Complex A - single cylinder}. In Fig.(\ref{modelsGSB} A) we present the model utilized for a single cylinder in the bchQRU triple mutant chlorosome. The cylinder comprises a stack of tightly-packed rings of Bchl$\,$\textit{c}  molecules. Each molecule is treated as a dipole with a squared dipole moment $|\vec \mu|^2 = 30 \ \mbox{D}^2$, possessing a well defined orientation and position in  space~\cite{gulli2019macroscopic}. Each ring has a radius   $R = 6 \ \mbox{nm} $ and contains $60 $ Bchl molecules separated by a nearest neighbour distance  $d = 0.628 \ \mbox{nm}$. The rings are equally separated by a vertical distance   $h = 0.83 \ \mbox{nm}$~\cite{gulli2019macroscopic, ganapathy2009alternating, gunther2016structure}. This choice of parameters for the geometry of the bchQRU triple mutant type is in agreement with Ref.~\cite{gunther2016structure}, where  the same cylindrical structure is obtained by wrapping the planar lattice under the rolling angle of $\delta=0^\circ$  with respect to the vertical axis of the lattice. In our simulations the single cylinder complex can contain up to \numprint{10800} Bchl\thinspace\textit{c} molecules, resulting in a maximum length  $L_{\textrm{max}} = 148.57 \ \mbox{nm}$, corresponding to a stack of $180$ rings. Further details about the geometry of the bchQRU triple mutant concerning the positions and orientations can be found in Ref.~\cite{gulli2019macroscopic}.

\textbf{Complex B - four concentric cylinders.} Chlorosomes of GSB display a  complex arrangement of molecules on a multi-wall structure ~\cite{gulli2019macroscopic, gunther2016structure, huh2014atomistic,linnanto2008investigation,saikin2014chromatic}. 
Panel B of Fig.~(\ref{modelsGSB}) depicts the model we developed, wherein Bchl~\textit{c} molecules form four concentric cylinders.
On each cylinder, the dipoles positions and orientations are consistent with those found in the single cylinder model of the bchQRU triple mutant, as explained in~\cite{gulli2019macroscopic}. The innermost cylinder has a radius $R = 3$\,\mbox{nm}, while the distance from wall to wall is $d = 2.1$~\mbox{nm}~\cite{oostergetel2010chlorosome, 
gulli2019macroscopic}.  

\textbf{Complex C - three adjacent concentric cylinders}. In panel C of Fig.~(\ref{modelsGSB}) we show a more elaborate model formed by three adjacent structures, each comprising four concentric cylinders. The wall-to-wall distance between adjacent cylinders is set to $3\,$\mbox{nm}, a realistic value according to Ref.~\cite{linnanto2013exciton}. The maximum length considered here is $L_{\textrm{max}}= 148.57\,$ \mbox{nm}, corresponding to $180$ vertical rings in each cylindrical structure. The largest system considered in our simulations  contains  \numprint{132840} Bchl\thinspace\textit{c} molecules, comparable with natural sizes.
The positions and orientations of dipoles on each tubular wall are the same as in Complex A and B.
    
\subsection{Antenna complexes in Purple bacteria (PB). }
\label{PBmodel-vesicle}
\begin{figure}[!ht]
    \centering
    \includegraphics[width=\columnwidth]{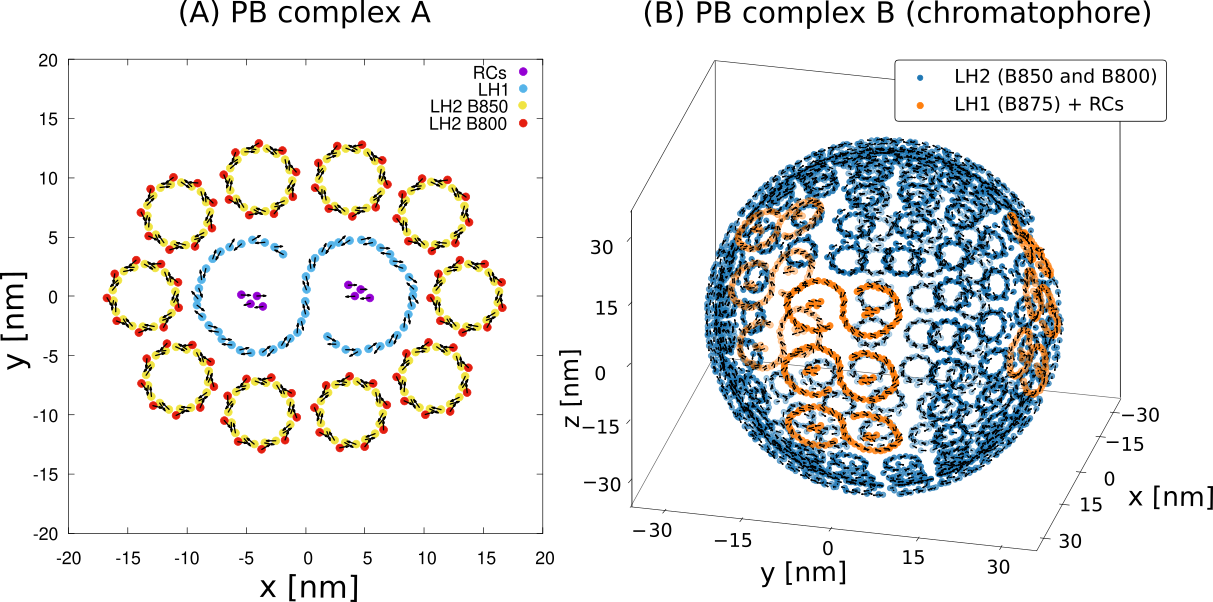}
    \caption{ {\it Architecture of PB light-harvesting complexes.} The two complexes A and B of Purple bacteria \textit{Rhodobacter sphaeroides} light-harvesting systems are shown. (A) Complex A: light-harvesting system  including two reaction centers RCs (purple) surrounded by the antenna complex LHI B875 (cyan), which is surrounded by 10 LHII B850 and LHII B800 (respectively  yellow and orange). Complex A light-harvesting system contains $N=334$ Bchl molecules.  Black arrows represent the positions and orientations of transition dipole moments associated to each Bchl\textit{a} molecule. 
    (B) Complex B: the vesicle with a radius $R=30$ nm contains $9$ LHI+RC complexes (orange S-shaped complexes) and $131$ LHII (B800 + B850) complexes (light blue circles). The total number of molecules is $N=\numprint{4113}$ and the area of the empty spherical cap at the south pole is $A_{emp} = 39.17$ $\mbox{nm}^2$ (for more details about the geometry  see~\cite{Note1}).
}
    \label{modelsPB}
\end{figure}

In  Purple bacteria \textit{Rhodobacter Sphaeroides}, the light-harvesting complexes  are organized in thousands of spherical membrane-embedded protrusions, called chromatophores which typically measure around $60$ nm in diameter.   Depending on the light conditions, Purple bacteria can contain from $500$  up to \numprint{2500}   chromatophores made of $\sim \numprint{5000}$ Bchl$\,$ molecules each~\cite{csener2007atomic}.
The major pigments in the chromatophores are the  Bchl$\,$\textit{a} ~\cite{qian2013three, koolhaas1998identification} which   are organized on different complexes inside the chromatophores.  
Here we consider a portion of the chromatophore (complex A) and the whole chromatophore (complex B):

\textbf{Complex A - light-harvesting system subunit.} The RC is directly surrounded by the so-called light harvesting
complex I (LHI). LHI is in turn surrounded by several smaller light harvesting complexes, the LHII complexes~\cite{damjanovic2000excitation,linnanto1999electronic,csener2007atomic}. As shown in panel A of Fig.~(\ref{modelsPB}) 
two RCs are surrounded by the light-harvesting complex LHI B875, which is an S-shaped
structure with 56 Bchl molecules. Then the LHI aggregate is surrounded by 10 LHII rings,
each composed by 2 units, the B800 ring, composed by 9 Bchl molecules, and the B850
ring with 18 Bchl molecules. Both complexes LHI and LHII are J-aggregates. In particular
LHI is a J-aggregate with two superradiant states close to the lowest excitonic state at 875
nm which are polarized in the ring plane. For the LHII aggregate, the B850 ring has two
superradiant states at 850 nm, while the B800 ring has a main absorption peak at 800 nm.
 This hierarchical structure is able to absorb photons at different
frequencies and to funnel the collected energy to the RC~\cite{mattiotti1, strumpfer2012excited, csener2007atomic, kassal2, kassal}.  In our simulation we model the RC  as  an aggregate of four
BChls:  two forming  the tightly-coupled special pair and two accessory
ones~\cite{kassal2}. Even if other molecules are present in the RC, these four Bchl molecules are the most relevant for the interaction with the electromagnetic field~\cite{michel1988relevance}.

\textbf{Complex B: the chromatophore.} We model the whole 3D structure of the chromatophore using  data from Ref.~\cite{strumpfer2012excited} collected by AFM, cryo-EM, X-ray crystallography and NMR measurements. In  panel B of Fig.~(\ref{modelsPB}) our model of the chromatophore is shown with  different substructures: i) LHII (blue rings); ii) LHI (orange structures); iii) RC, placed at the center of the LHI structures. Chromatophores, see Fig.~(\ref{modelsPB} B), are connected to the cell membrane at their south pole, therefore the southern polar region is left empty (no molecular aggregates are placed there) to allow the contact with the cell membrane~\cite{Note1}. The chromatophore spherical vesicles are reproduced by mapping small planar regions onto spherical ones containing the LHI-RCs or an LHII complex with the area-preserving inverse-Mollweide transformations, see Fig.~(\ref{PBprojection}) in~\autoref{AppA} of the {\it Supporting Information}, used also in Ref.~\cite{strumpfer2012excited,lapaine2011mollweide,snyder1987map}. While in nature the ratio between LHII and LHI-RC depends on illumination and other conditions, here we consider a vesicle model that contains  $9$ S-shaped LHI-RCs and $131$ LHII complexes. The overall Bchl/RC ratio used in our simulation ($229$) is within the natural range for these vesicles, which  is between $108$ and $248$ and corresponds to low-light growth conditions~\cite{strumpfer2012excited} where these systems show the largest efficiency.  The data for the LHI, LHII and RC are taken from Ref.~\cite{kassal2,csener2007atomic}. Even if it would be more realistic to add random rotations of LHII rings around the axis joining the center of the vesicle and the center of the LHII rings, we checked that the effect of such random rotations is small, see  Fig.~(\ref{PBrotationLHII}) of~\autoref{AppE}, and thus we did not include such rotations in our simulations. 

\subsection{The radiative Hamiltonian and the dipole approximation}

Since photosynthetic antennae operate under natural sunlight, which is very dilute, the single-excitation approximation can be used, so that only states containing a single excitation have been considered. 
Choosing the basis states in  the single-excitation manifold, where $|i\rangle$ represents  a state in which the $i^{th}$ molecule is excited while all the others are in the ground state, the systems can be described 
 through a radiative non-Hermitian Hamiltonian which takes into account the interaction between the molecules mediated by the electromagnetic field
(EMF)~\cite{mukamel,kaiser,grad1988radiative,gross1982superradiance}.  The radiative Hamiltonian reads:     
\begin{equation} 
  H=\sum_{i=1}^N e_0|i\rangle \langle i|+\sum_{i\neq j}\Delta_{ij}|i\rangle \langle j|-\frac{\i}{2}\sum_{i,j=1}^{N}Q_{ij}|i\rangle \langle j| \, .
  \label{eq:ham}
\end{equation} 
where $e_0$ is the excitation energy of single emitter (Bchl molecule in our case). The terms $\Delta_{ij}$ and $Q_{ij}$ have a diagonal part  given  by: 
\begin{equation}
  \Delta_{jj} = 0 \, , \qquad
  Q_{jj} = \frac{4}{3} \mu^2 k_0^3 = \gamma \, , \label{eq:gamma}
\end{equation}
with $\mu=|\vec{\mu}|$ being the transition dipole moment (TDM) and $k_0=\frac{2 \pi}{ \lambda_0}$, where $\lambda_0$ is the wavelength associated with the molecular transition. The off-diagonal part ($i \ne j$) is given  by
\begin{align}
  \Delta_{ij} = &\frac{3\gamma}{4} \left[ \left( -\frac{\cos (k_0 r_{ij})}{(k_0 r_{ij})} +
    \frac{\sin (k_0 r_{ij})}{(k_0 r_{ij})^2} + \frac{\cos (k_0 r_{ij})}{(k_0 r_{ij})^3} \right)
    \hat{\mu}_i \cdot \hat{\mu}_j\right. \nonumber \\
    &-\left. \left( -\frac{\cos (k_0 r_{ij})}{(k_0 r_{ij})} + 3\frac{\sin (k_0 r_{ij})}{(k_0 r_{ij})^2} +
    3\frac{\cos (k_0 r_{ij})}{(k_0 r_{ij})^3}\right) \left( \hat{\mu}_i \cdot \hat{r}_{ij}
    \right) \left( \hat{\mu}_j \cdot \hat{r}_{ij} \right) \right],
    \label{eq:d1}
\end{align}
\begin{align}    
  Q_{ij} = &\frac{3\gamma}{2} \left[ \left( \frac{\sin (k_0 r_{ij})}{(k_0 r_{ij})} +
    \frac{\cos (k_0 r_{ij})}{(k_0 r_{ij})^2} - \frac{\sin (k_0 r_{ij})}{(k_0 r_{ij})^3} \right)
    \hat{\mu}_i \cdot \hat{\mu}_j\right. \nonumber \\
  &-\left. \left( \frac{\sin (k_0 r_{ij})}{(k_0 r_{ij})} + 3\frac{\cos (k_0 r_{ij})}{(k_0 r_{ij})^2} -
    3\frac{\sin (k_0 r_{ij})}{(k_0 r_{ij})^3}\right) \left( \hat{\mu}_i \cdot \hat{r}_{ij}
    \right) \left( \hat{\mu}_j \cdot \hat{r}_{ij} \right) \right], 
    \label{eq:g1}
\end{align}
where $\hat{\mu}_i :=  \vec{\mu}_i  /
\mu$ is the unit dipole moment of the $i^{th}$ site and $\hat{r}_{ij} := \vec{r}_{ij}
/ r_{ij}$ is the unit vector joining the $i^{th}$ and the $j^{th}$ sites. See~\autoref{AppB} of the {\it Supporting Information} for the parameters we used for GSB and PB in the Hamiltonian.

Diagonalizing  the Hamiltonian~(\ref{eq:ham}) we obtain the  complex eigenvalues $
\varepsilon_{n}=\mbox{E}_n-\i\frac{\Gamma_{n}}{2}$
where $\Gamma_{n}$ is the radiative decay of the $n^{th}$ eigenstate. In general $\Gamma_{n}$ differs from the radiative decay of the single molecule $\gamma$.   In particular, when the ratio $\Gamma_{n}/\gamma \gg 1$   we will talk about a ``superradiant state'' (SRS) or bright state, otherwise when $\Gamma_n/\gamma \ll 1$ the  state is called ``subradiant'' or dark. In other words, a SRS  can radiate much faster than a single molecule, while a subradiant  one radiates at a rate much slower than the single molecule radiative decay. \\

If  the non-Hermitian term $Q_{ij}$ can be considered a small perturbation,
the optical absorption of an eigenstate  can be  estimated in terms of its dipole strength, computed using only  the Hermitian part of the Hamiltonian~(\ref{eq:ham}). 
Denoting the $n^{th}$ eigenstate of the Hermitian part of the Hamiltonian~(\ref{eq:ham}) with $|E_n\rangle$, we can expand it on the site basis, so that 
\begin{equation} \label{eq:expan}
|E_{n}\rangle=\sum_{i=1}^{N} C_{ni} \, |i\rangle.
\end{equation}
To each basis state $|i\rangle$, a dipole moment $\vec{\mu}_i$ is associated, corresponding to the TDM of the $i^{th}$ molecule. 
If $N$ is the total number of molecules, then we will express the TDM $\vec{D}_n$ associated with the $n^{th}$ eigenstate as follows: 
\begin{equation} \label{eq:dipst} 
\vec{D}_n=\sum_{i=1}^{N} C_{ni} \, \hat{\mu}_i. 
\end{equation} 
The dipole strength of the $n^{th}$ eigenstate is defined  by $|\vec{D}_n|^{2}$ (note that   due to normalization   $\sum_{n=1}^{N} |\vec{D}_n|^{2}=N$).

The non-Hermitian part of the radiative Hamiltonian~(\ref{eq:ham}) can be treated as a perturbation whenever the decay widths are much smaller than the mean level spacing computed using the real part of the complex eigenvalues. When this criterion, known as \textit{resonance overlap criterion}~\cite{zelevinsky}, is valid, one can exclusively utilize the Hermitian part of the Hamiltonian. This reduction in complexity accelerates calculations. The Hermitian part of the Hamiltonian~(\ref{eq:ham}) is defined as follows:
\begin{equation} \label{eq:hreal}
H_{H}=\sum_{i=1}^N e_0|i\rangle \langle i|+\sum_{i\neq j}\Delta_{ij}|i\rangle \langle j|,
\end{equation}
where $\Delta_{i,j}$  is given in Eq.~(\ref{eq:d1}).
Note that when the resonances do not overlap we have   $|\vec{D}_n|^2 \approx \Gamma_n/\gamma$.

 Finally, we note that when resonances do not overlap and  the system size is much smaller than $\lambda_0$  (i.e. when $k_{0}r_{ij}\ll 1$), the Hermitian part of the radiative Hamiltonian reduces to  the standard dipole-dipole Frenkel Hamiltonian: 

\begin{equation}
    \label{dip}
H_{dip}=\sum_{i=1}^N e_0|i\rangle \langle i|+\sum_{i\neq j}\displaystyle \frac{\vec{\mu}_{i} \cdot \vec{\mu}_{j}-3(\vec{\mu}_{i} \cdot \hat{r}_{ij})(\vec{\mu}_{j} \cdot \hat{r}_{ij})}{r_{ij}^{3}}|i\rangle \langle j|, 
\end{equation}

In the following,    we will  analyzed all complexes using the three different Hamiltonian models introduced in this section, namely: 
\begin{enumerate}
\item NHH: non-Hermitian radiative Hamiltonian Eq.~(\ref{eq:ham}). 
\item HH: Hermitian Hamiltonian Eq.~(\ref{eq:hreal}) valid under the non-overlapping resonance criterion.  
\item DH: Dipole Hamiltonian Eq.~(\ref{dip}) valid under the non-overlapping resonance criterion and when the system size is small compared to the wavelength associated with the optical transition of the molecules. 
\end{enumerate}

\subsection{Robustness to thermal noise and static disorder}

Modelling the effect of noise and disorder in natural systems has been an important issue in the field of quantum biology. The environmental noise in the literature  has been described with an open quantum system approach using an environment-system interaction  even for non-Markovian baths~\cite{caruso2010entanglement,ishizaki2009theoretical}. Nevertheless due to the huge sizes considered in this manuscript we will adopt a minimal but solid modelling of noise and disorder. Noise due to thermal fluctuations will be taken into account by considering a state at thermal equilibrium, while disorder will be modelled as  space dependent and time independent fluctuations of the site energies, keeping the couplings between the molecules constant. Both approaches have been widely used in literature~\cite{celardo2014cooperative,monshouwer1997superradiance,mukamel,schulten1:1}. They have the advantage to capture the main detrimental effects of noise and disorder even if a realistic modelling of the environment would require a much more sophisticated approach~\cite{ref1,ref2,ref3,ref4,ref5}. 

Given a quantum state specified by the density matrix $\hat{\rho}$ it is possible to define its coherence length in the single-excitation manifold defined by the  basis states $\ket{i}$~\cite{damjanovic2002excitons,strumpfer2009light,gulli2019macroscopic} in the following way: 
\begin{equation} \label{eq:lrho}
L_{\rho}=\frac{1}{N}\frac{\left(\sum_{ij}|\rho_{ij}|\right)^2}{\sum_{ij}|\rho_{ij}|^2}.
\end{equation}
$L_{\rho}$  in Eq.~(\ref{eq:lrho})
measures how much a single excitation is spread coherently over the molecules composing the aggregate. 
To give an idea of its physical meaning let us consider three different simple cases: 
\begin{itemize}
\item   a pure localized state,  $\hat{\rho}=|i \rangle\langle i|$; then it is easy to see that the  coherence length defined in Eq.~(\ref{eq:lrho}) is given by $L_{\rho}=1/N$.
This case represents the minimal value that
$L_\rho$ can get.
\item   A completely delocalized mixed state  characterized by the density matrix

\begin{equation}
\mbox{$\hat{\rho}=(1/N) \sum_{i=1}^{N} |i\rangle\langle i|$}. 
\label{eq:rho}
\end{equation}
In this case we have  $L_{\rho}=1$. This state is maximally delocalized in the basis, but it is completely incoherent. 
\item  Lastly we consider the fully delocalized  coherent state: $\hat{\rho}=(1/N) \sum_{i,j=1}^{N} |i\rangle\langle j|$.
In this case we have  $L_{\rho}=N$. Note that any pure state with constant amplitude $1/\sqrt{N}$ over the sites and arbitrary phases would give the same result.
\end{itemize}
It is easy to see that $1/N \leq L_{\rho} \leq N$.
The closer $L_{\rho}$ is to $N$, the higher a coherent delocalization  can be assigned to our state.
In the same way $L_\rho < 1 $ indicates an incoherent localized state. States characterized by $L_\rho \sim 1 $ have a little ambiguity (since both localization  and coherence are  measured on the same length scale).

For all  models   we have computed  the thermal coherence length at room temperature ($T=300$~K), defined for a state at the canonical equilibrium and  whose matrix elements are given by:
\begin{equation} \label{eq:expand}
\rho_{ij}=\sum_{n} \frac{e^{-\beta E_n}}{\mbox{Tr}({e^{-\beta \hat{H}}})} \langle i|E_n\rangle \langle E_n|j\rangle,
\end{equation}
where $\beta=1/k_B T$. 

A very important question to be answered is how much the symmetrical arrangement of the molecules that produces superradiance,
  is also able to produce a large thermal coherence length  at room temperature. Note that even if we consider the coherence length at thermal equilibrium, this does not mean that out-of-equilibrium processes are not important in molecular nanotubes. Indeed  in Ref.~\cite{dostal2012two} strong evidence of ultra-fast transport in  natural structures with transfer times less than 100~fs have been discussed. Nevertheless thermal equilibrium can be considered as a worst case scenario for coherences. For this reason assuming thermal equilibrium can be considered a good starting point to assess the structural robustness of quantum coherence to thermal noise. 
In the following we calculate the coherence length $L_{\rho}$ according to Eq.~(\ref{eq:lrho}), using a thermal density matrix as in Eq.~(\ref{eq:expand}).\\

Natural complexes are not only affected by thermal noise but also by other sources of disorder due to the fluctuations in the local environment. 
 In order to analyze the robustness to this kind of disorder, we have considered time-independent and space-dependent fluctuations of the excitation energy of the molecules in the aggregates. Specifically, we consider energy fluctuations which are uniformly distributed around the excitation energy of the molecules $e_0$, between $e_0 - W/2$ and $e_0 + W/2$, where $W$ represents the disorder strength. It is known that static disorder induces eigenmodes localization~\cite{anderson1958absence} and quenching of superradiance~\cite{celardo2014cooperative}.  

\section{Numerical results}
\label{numerical}
In this section we show the  numerical results obtained for the  antenna complexes of GSB and PB. 
Note that in~\autoref{Spectra}  of \textit{Supporting Information} the absorption  spectra derived using our model are compared with the experimental results for GBS and PB, showing a good agreement.

\subsection{Green Sulfur Bacteria complexes}
\label{numericalGSB}
\begin{figure}[!ht]
    \centering
    \includegraphics[width=\columnwidth]{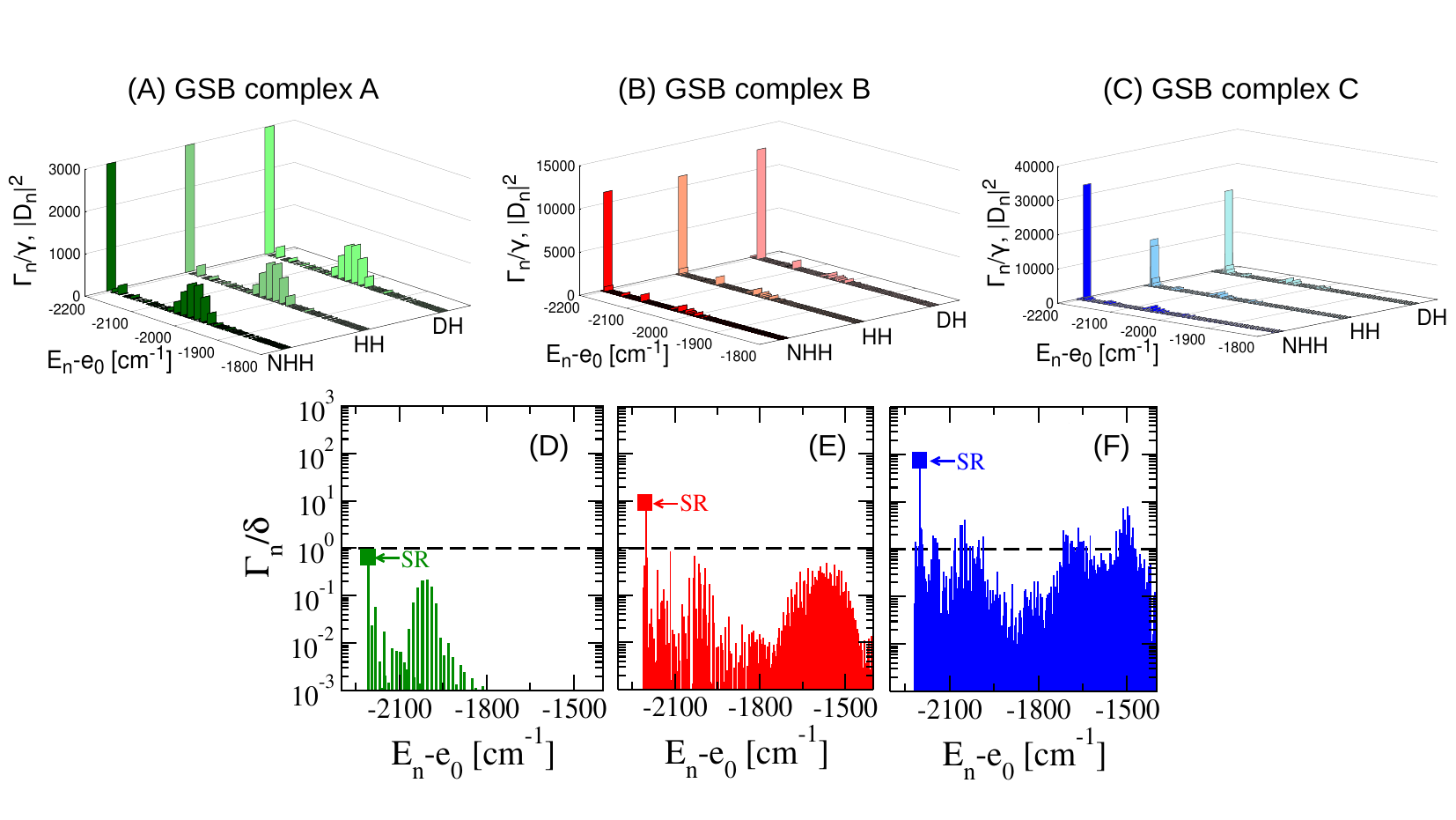}
    \caption{{\it Superradiant states in Green Sulfur Bacteria Antennae.}  Complexes A (single cylinder, green), B (four concentric cylinders, red) and C (three adjacent concentric cylinders, blue) are shown in the respective panels (Panels A,B,C). Three different Hamiltonian models are compared for each complex using different shades of colour (NHH: non-Hermitian Hamiltonian Eq.~(\ref{eq:ham}, dark), HH: Hermitian Hamiltonian Eq.~(\ref{eq:hreal}, medium shade), DH: Dipole Hamiltonian Eq.~(\ref{dip}, light). The squared dipole strength $|\mbox{D}_n|^2$, see Eq.~(\ref{eq:dipst}), is shown for the HH and DH model, while the radiative decay width $\Gamma_n/\gamma$ is shown for the NHH model  as a function of the energy $E_n-e_0$. Panels (A-C) show only the lowest part of the energy spectrum where the most superradiant states are located. Results are computed by using a fixed length $L= 148.57 $ nm for each aggregate, which corresponds to the maximal length analyzed (180 rings for each cylinder). (Panel D-F) Ratio between the decay width $\Gamma_n$ obtained by diagonalizing the full radiative Hamiltonian (NHH model) in Eq.~(\ref{eq:ham}) and the mean level spacing $\delta$ as a function of the energy $E_n-e_0$ for complexes A (dark green), B (dark red), C (dark blue) with a fixed length $L= 148.57 $ nm. The mean level spacing $\delta$ is computed as the ratio between the energy spectral width and the total number of eigenmodes for each complex.  Green, blue and red squares indicate the positions in the energy spectra of the most superradiant state for each complex. 
    The horizontal dashed line
    represents the value of the ratio ($\Gamma_n/\delta=1$) above which resonances overlap.}
    \label{f1}
\end{figure}

\begin{figure}[!ht]
    \centering
    \includegraphics[scale=0.61]{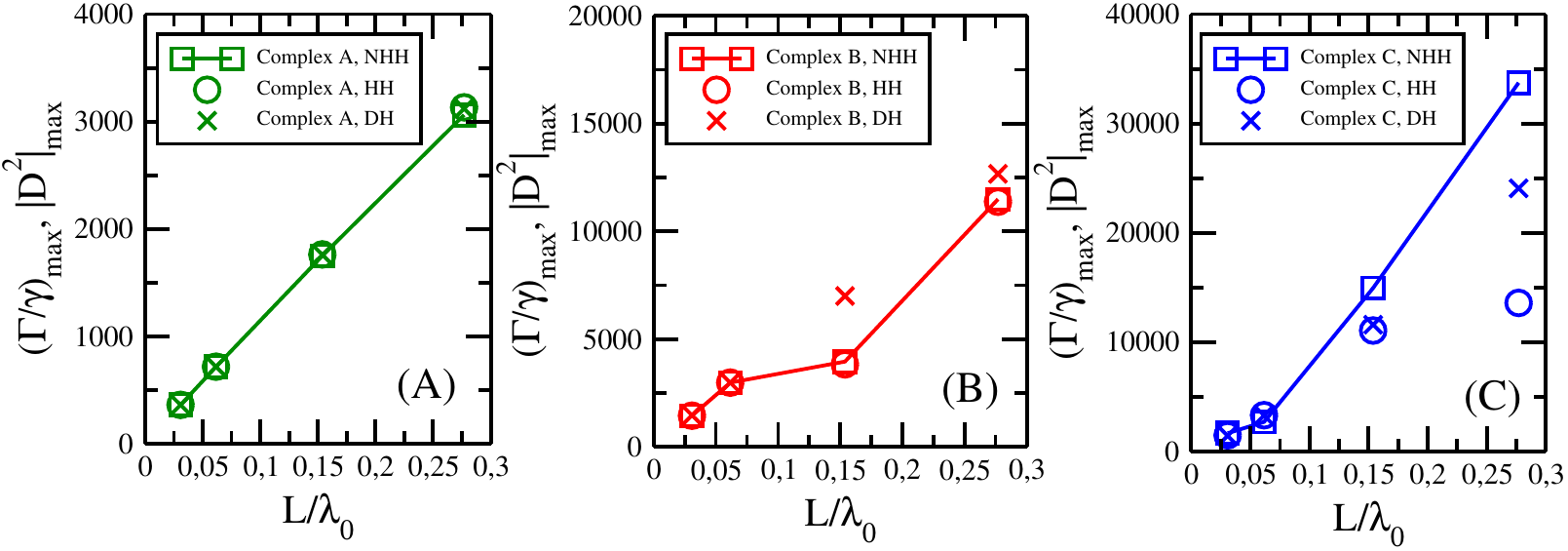}
    \caption{{\it Cooperativity in GSB complexes.} Panels A, B, C: the maximal decay width is shown for different complexes A (green symbols in panel A), B (red symbols in panel B), C (blue symbols in panel C) as a function of the length of the aggregates $L$, normalized to the transition wavelength of a single molecule $\lambda_0$, see Tab.~(\ref{tabGSB}) in~\autoref{AppB}.  The maximal decay width has been obtained from the dipole strength $|D|^2_{max}$ for the DH model (crosses) and the HH model (circles) and from $(\Gamma/\gamma)_{max}$ for the NHH model (squares). For the NHH model, which is the most accurate model, symbols (squared) have been connected  by lines. 
    The maximal length we considered is $L=148.57$ nm, that corresponds to aggregates made of cylinders with $180 $ rings.
}
    \label{f2}
\end{figure}

Here, the  complexes A, B, C of GSB  described in  Sec.~\ref{subsec:GSBmodel} 
have been analyzed. In Fig.~(\ref{f1}) the spectrum of the three complexes is shown for the maximal length considered, $L=148.57$ nm, corresponding to cylindrical aggregates with  180 rings. Thus, for the three complexes  we have  $N=\numprint{10800} $ (complex A), $N=\numprint{44280}$ (complex B), $N=\numprint{132840}$ (complex C) chlorophyll molecules.  Panels (A-C) in Fig.~(\ref{f1}) show that the structure of all complexes allow for the emergence of red-shifted superradiant states, close to the lowest excitonic states. Moreover the amount of maximal  superradiance ($(\Gamma/\gamma)_{max}$) increases with the system size, see Fig.~(\ref{f2}). The largest amount of  superradiance is present in complex C which is the largest one and the closest to the natural size and structure of GSB antennae. Note that the fact that in larger aggregates superradiance is enhanced  is far from being trivial. Indeed, if we would increase the system size by adding the molecules in the same positions but with randomized dipole directions, no superradiant enhancement would be present, as shown in~\autoref{AppF-GSB} of the {\it Supporting Information}.

By comparing the three Hamiltonians, see also discussion in~\autoref{AppC}, we note that for complex A, all of them give a good description of the superradiant states. For complex B a deviation between DH and  the other two models (NHH and HH) is observed, while in complex C the three Hamiltonian models give different results. This shows that using the most accurate Hamiltonian, which is the NHH model, is essential to describe large photosynthetic antennae. The same behavior is also shown in Fig.~(\ref{f2}). In each panel of Fig.~(\ref{f2}) we show the maximal decay width for each complex as a function the complex  length. In each panel the maximal decay width has been computed using 
 three different Hamiltonian models: DH, HH, NHH. For the DH and HH model we computed the maximal decay width using the dipole strength, while for the NHH model the maximal decay width has been computed  using the imaginary part of the complex eigenvalues of the radiative non-Hermitian Hamiltonian.  As one can see in the largest complex Fig.~(\ref{f2} panel C) the three Hamiltonians give very different results. 
 
The differences between the NHH and the other Hamiltonian models is not due to the coupling strengths, which are almost identical in the three Hamiltonian models, as shown in Tab.~(\autoref{tab-coupling}) in~\autoref{average-coupling}, but it can be explained by the overlapping resonance criterion~\cite{zelevinsky}, see panel D in Fig.~(\ref{f1}).
 Indeed, when the decay width becomes comparable with the eigenmodes mean level spacing, the non-Hermitian part of the Hamiltonian cannot be treated perturbatively and the dipole strength~Eq.~(\ref{eq:dipst}) does not describe anymore the decay widths of the eigenmodes.

In order to 
analyze the robustness of  natural models to static disorder and thermal noise we have chosen two figures of merit: the maximal dipole strength Eq.~(\ref{eq:dipst}) and the thermal coherence length Eq.~(\ref{eq:lrho}).  Both quantities  have been studied as a function of the static disorder strength $W$.

\begin{figure}
    \centering
    \includegraphics[width=\columnwidth]{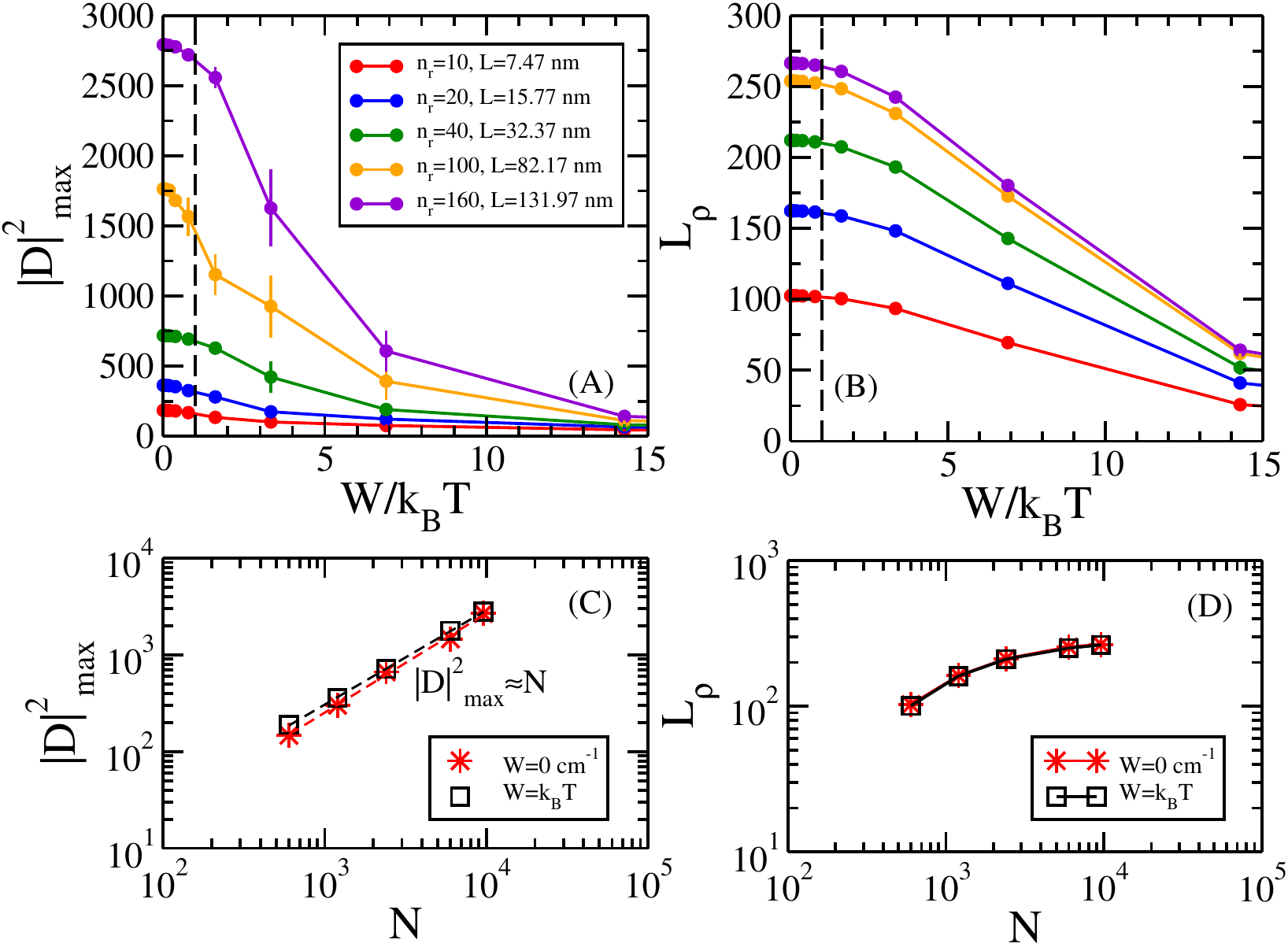}

    \caption{{\it GSB complex A: robustness to disorder and thermal noise.} The average maximal dipole strength (panel A) defined in Eq.~(\ref{eq:dipst}) and the thermal coherence length (panel B) shown in Eq.~(\ref{eq:lrho}) as a function of the normalized static disorder $W/k_B T$ are shown for a GSB  single wall cylinder (complex A)  with different lengths $L$, comprising $n_r$ rings. The thermal coherence length and the maximal dipole strength are computed using the HH model Eq.~(\ref{eq:hreal}) and averaging over $10$ disorder realizations. $k_B T$ is the thermal energy at room temperature ($T=300 \ \mbox{K}$). The vertical dashed line represents  $W=k_BT$. Panels (C-D): average maximal dipole strength (panel C) and thermal coherence length (panel D) as a function of the number $N$ of Bchl molecules at zero disorder (red stars) and at room temperature (black squares).   }
    \label{f3}
\end{figure}

Fig.~(\ref{f3} A,B) shows the  maximal dipole strength and the thermal coherence length for the case of a single cylinder (complex A of the GSB) for different cylinder lengths. For complex A we used  the HH model to compute both figures of merits since the results obtained with the latter model do not differ substantially from those obtained with the NHH model  while giving  a substantial computational advantage. 

In Fig.~(\ref{f3} A,B), one can see that both quantities increase with the system size, even if the thermal coherence length tends to saturate. Moreover, both figures of merits show that single cylinders are very robust to disorder and thermal noise. Indeed, their maximal value (for $W=0$) remains mostly unchanged up to values of the disorder strength comparable with the 
thermal energy at room temperature ($W=k_BT$ for $T=300$ K). Note that, in natural systems,  the static disorder strength is usually of the same order of magnitude of the thermal energy. 
In Fig.~(\ref{f3} C,D) the maximal dipole strength (panel C) and the  thermal coherence length (panel D) are shown at $W=0$ and $W=k_BT$  for different cylinder length containing different number $N$ of chlorophyll molecules. Interestingly the value of both figure of merits  is the same for the two values of disorder strength considered, showing their extreme robustness to disorder.  Moreover the maximal dipole strength (panel C) increases linearly with $N$ which shows that such structures are extremely effective at preserving cooperative effects as a function of the system size. Indeed single excitation superradiance cannot increase faster than $N$. The thermal coherence also increases with the system size (panel D), even if it shows a tendency to saturate as $N$ increases.

\begin{figure}
    \centering
    \includegraphics[width=\columnwidth]{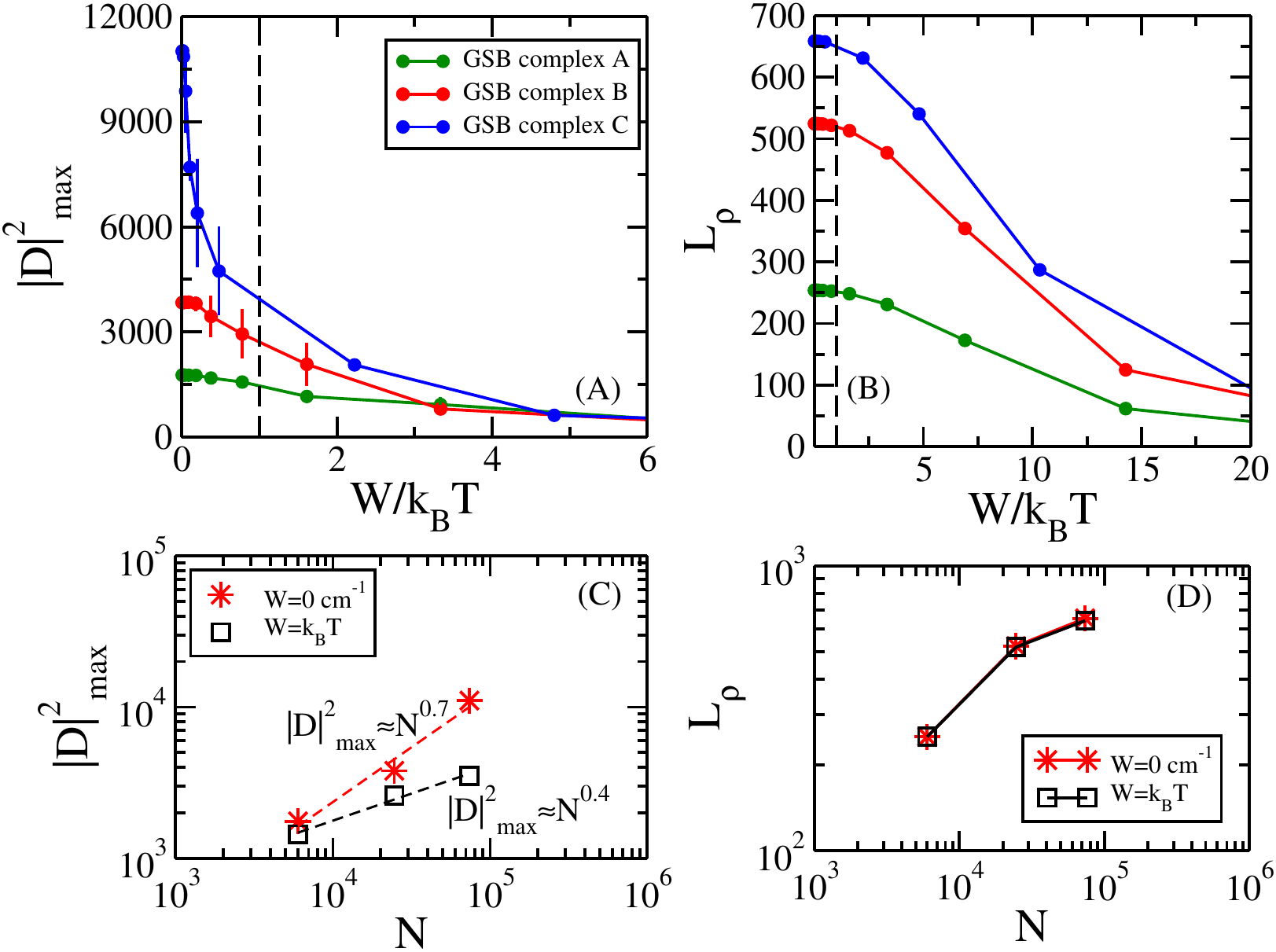}

    \caption{{\it Robustness to static disorder and thermal noise: comparison between GSB complexes A,B,C.} Comparison between aggregates with a single cylinder (complex A,  green circles), four concentric rolls (complex B,  red circles)  and three adjacent aggregates (complex C,  blue circles). The average thermal coherence length and the average maximal dipole strength have been  computed using the HH model Eq.~(\ref{eq:hreal}) and averaging over $10$ disorder realizations for complexes A and B and $5$ disorder realization for complex C.  (Panel A) Average maximal dipole strength defined in Eq.~(\ref{eq:dipst})  as a function of  of the normalized static disorder. (Panel B) Average thermal coherence length, Eq.~(\ref{eq:lrho}),   as a function of the normalized static disorder.
    In both panels the complexes  have  the same length $L=82.17$ nm that corresponds to $n_r=100$ rings for each cylindrical aggregate.
     $k_B T$ is the thermal energy at room temperature $T=300 \ \mbox{K}$. The vertical dashed line represents  $W=k_BT$. Panels (C-D): average maximal dipole strength (panel C) and thermal coherence length (panel D) as a function of the number $N$ of Bchl molecules at zero disorder (red stars) and at room temperature (black squares).}
    \label{f4}
\end{figure}

We now consider in Fig.~(\ref{f4}), the three different complexes for a fixed length corresponding  to 100 rings for each cylindrical aggregates ($82.17$ nm),  smaller than the maximal system length considered in Fig.~(\ref{f1},\ref{f2}). Also in this case we used the HH model to compute both figures of merit since for this length scale the HH model is quite accurate for all complexes, see discussion in~\autoref{AppC} of the {\it Supporting Information}. While the maximal dipole strength decreases with disorder Fig.~(\ref{f4} A), it is still much larger than one even for $W=k_BT$. On the other hand, the thermal coherence length shows a large  robustness to disorder, Fig.~(\ref{f4} B) up to $W=k_BT$. 
In Fig.~(\ref{f4} C,D) the maximal dipole strength (panel C) and the  thermal coherence length (panel D) are shown at $W=0$ and $W=k_BT$ as a function of the number $N$ of chlorophyll molecules present in the three different complexes A,B,C. 
The most important feature of both figures of merit is that they grow as a larger portion of the photosynthetic antenna is considered, showing that the structure of the GSB photosynthetic antenna as a whole is able to support excitonic coherences. Note that in presence of disorder, the maximal dipole strength grows slower with $N$ compared with the case of zero disorder (panel C). On the hand,  the value of the coherence length is the same for the two values of disorder strength considered, showing its  robustness to disorder (panel D), but as $N$ increases it shows a tendency to saturate.

\subsection{Purple bacteria Complexes}

Here we analyze the whole chromatophore of the Purple bacteria antenna complex (complex B), see Fig.~(\ref{modelsPB} B) containing \numprint{4113} chlorophyll molecules.  As a  comparison we will also analyze complex A, see  Fig.~(\ref{modelsPB} A).

\begin{figure}
    \centering
    \includegraphics[width=\columnwidth]{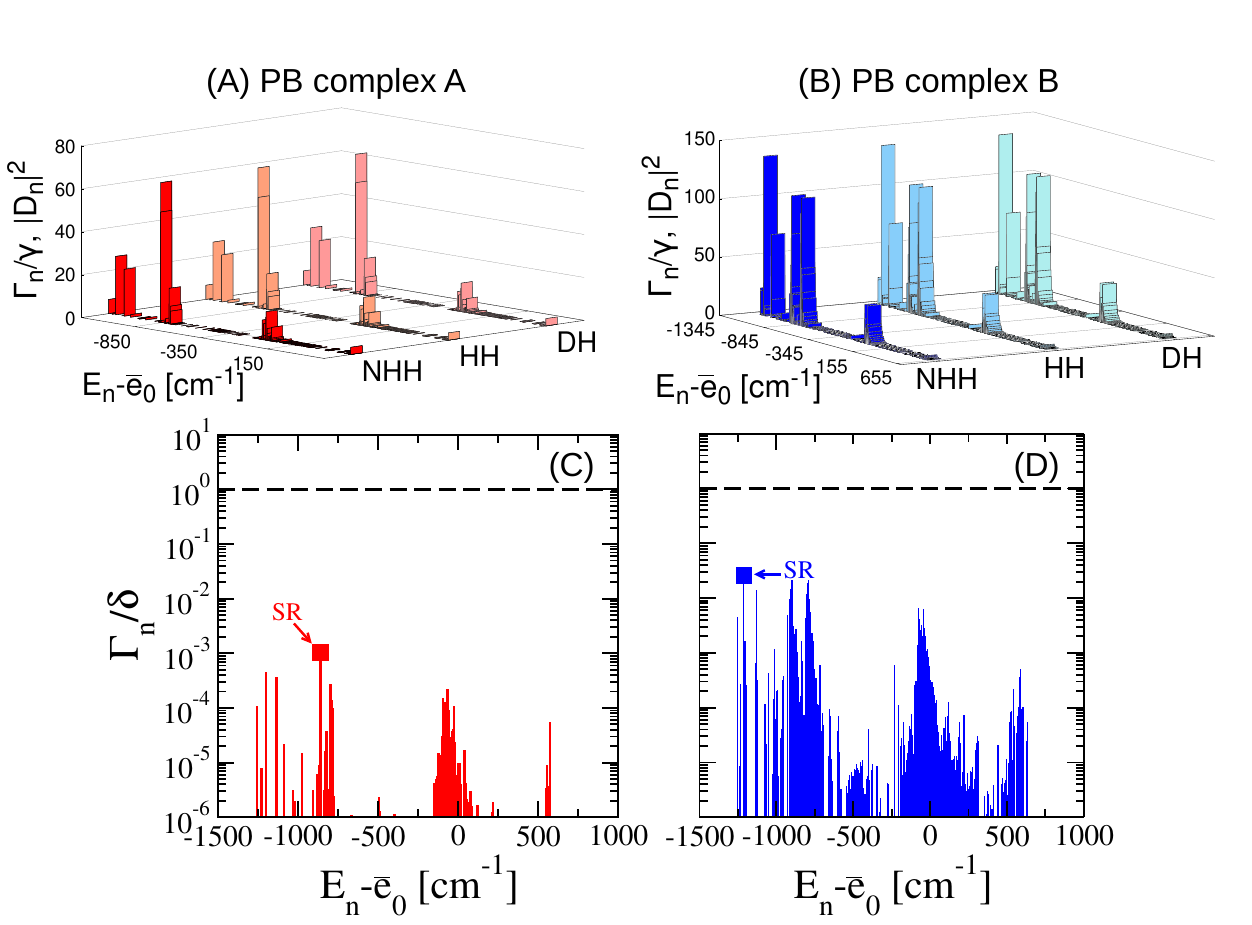}
    \caption{{\it Superradiant states and cooperativity in Purple bacteria Antennae.} 
    (Panels A) Complex A (light-harvesting system with an LHI S-shaped system with 2 RCs surrounded by 10 LHII rings in red colour). (Panels B) Complex B (entire PB chromatophore  in blue colour). Three different Hamiltonian models are compared for each complex using different shades of colour (NHH: non-Hermitian Hamiltonian Eq.~(\ref{eq:ham}, in dark colours), HH: Hermitian Hamiltonian Eq.~(\ref{eq:hreal}, in medium shade of colours), DH: Dipole Hamiltonian Eq.~(\ref{dip}, in light colours). In panels A-B, the squared dipole strength $|\mbox{D}_n|^2$, see Eq.~(\ref{eq:dipst}), is shown for the HH and DH models, while the radiative decay width $\Gamma_n/\gamma$ is shown for the NHH model  as a function of the energy $E_n-\overline{e}_0$, where $\overline{e}_0$ is the average excitation energy of the Bchl molecules found in PB antennae complexes, see Tab.~(\ref{tabPB}) in~\autoref{AppB}. 
    Panels (A-B) show only the lowest part of the energy spectrum where the most superradiant states are located.  Panels (C-D) show in log-scale the ratio between the decay width $\Gamma_n$, obtained by diagonalizing the full radiative Hamiltonian (NHH model) in Eq.~(\ref{eq:ham}), and the mean level spacing $\delta$ as a function of the energy $E_n-\overline{e}_0$ for complexes A (in dark red color) and B (in dark blue colour). The mean level spacing $\delta$ is computed as the ratio between the energy spectral width and the total number of eigenmodes for each complex.  Red and blue squares indicate the positions in the energy spectra of the most superradiant state for each complex.     The horizontal dashed line
    represents the value $\Gamma_n/\delta=1$ above which resonances overlap. }
    \label{PB3}
\end{figure}

Fig.~(\ref{PB3}) shows the spectra of the two complexes obtained by diagonalizing the three   Hamiltonians models (HH, NHH and DH). 
Panels (A-B) of Fig.~(\ref{PB3}) demonstrate that both PB complexes allow the emergence  of red-shifted superradiant states. Furthermore, the amount of superradiance is larger in the larger complex, the   chromatophore.  It is interesting to note that all Hamiltonian  models (HH, NHH and DH) give very similar results in both the complexes considered. The HH model and the NHH model give  similar results since, as it is shown in panel C of Fig.~(\ref{PB3}),  the decay widths are always smaller than the eigenmodes mean level spacing, so that the non-Hermitian part of the Hamiltonian can be treated perturbatively.   The DH model is also a good approximation even for the whole chromatophore since $L/\lambda \approx 0.1$ so that we are in the small volume limit.  

Finally, we analyze the robustness to static disorder and thermal noise using the average maximal dipole strength~(\ref{eq:dipst}) and the average thermal coherence length~(\ref{eq:lrho}) as figures of merit. Both of them have been computed using the HH model.

\begin{figure}[!ht]
    \centering
    \includegraphics[width=\columnwidth]{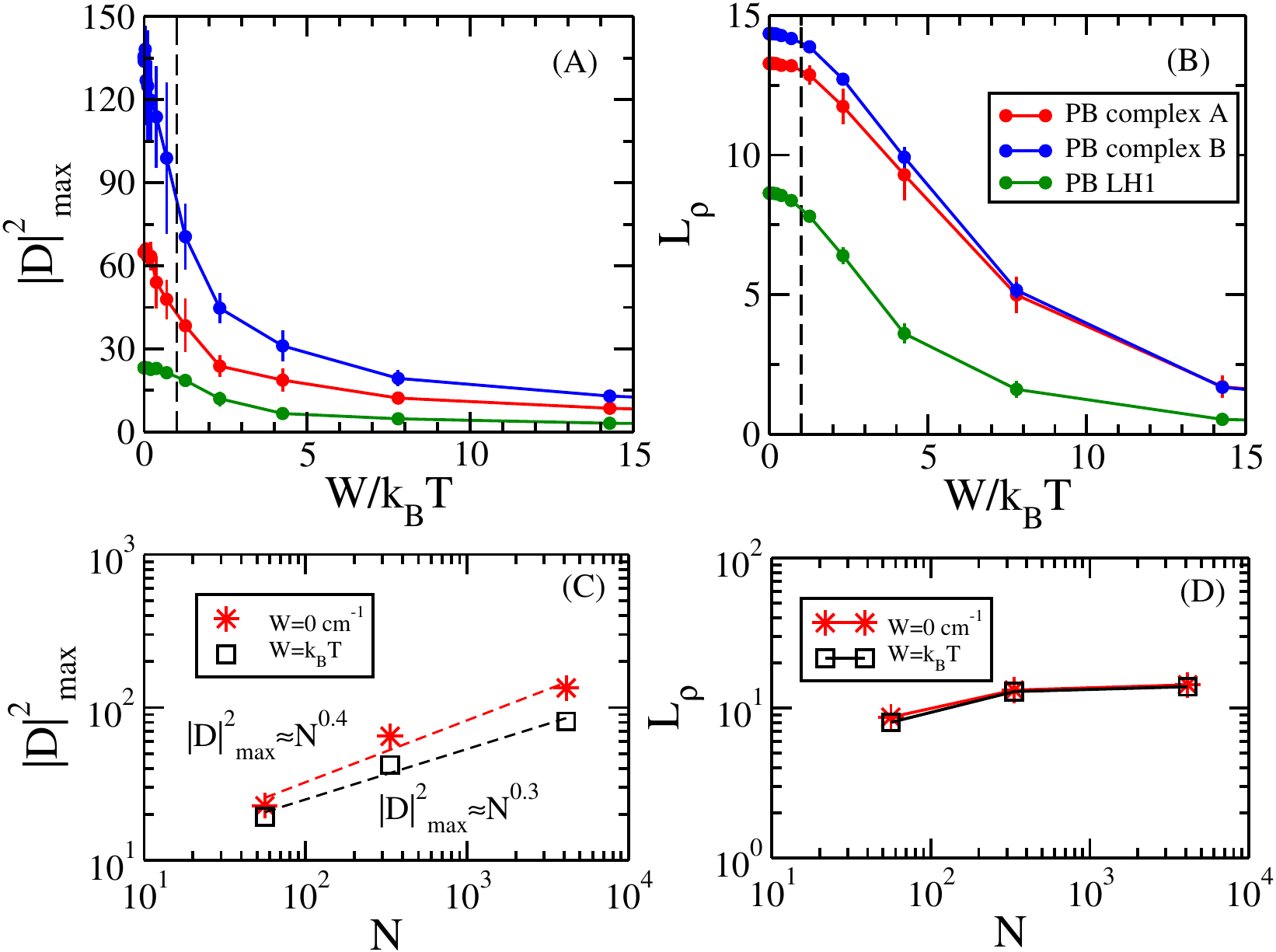}
    \caption{{\it Robustness to static disorder and thermal noise: comparison between PB complexes A,B and LHI system.} The average maximal dipole strength (panel A) and the average thermal coherence length  (panel B) are shown as a function of the normalized static disorder $W/k_BT$ for complexes A (red circles), B (blue circles) and a LHI S-shaped system with $N=56$ (green circles). $k_B T$ is the thermal energy at room temperature  $T=300 \ \mbox{K}$. The vertical dashed line represents  $W=k_BT$. The average thermal coherence length  and the maximal dipole strength  defined respectively in Eq.~(\ref{eq:lrho}) and Eq.~(\ref{eq:dipst}) have been computed by using the HH model Eq.~(\ref{eq:hreal}). For each value of the disorder strength we average the thermal coherence length and the maximal squared dipole strength over 10 disorder realizations. Panels (C-D): average maximal dipole strength (panel C) and thermal coherence length (panel D) as a function of the number $N$ of Bchl molecules at zero disorder (red stars) and at room temperature (black squares).  }
    \label{PB7}
\end{figure}

 In Fig.~(\ref{PB7})  the average maximal dipole strength and thermal coherence length  for both complexes A and B are shown as a function of the disorder strength $W$ rescaled over the thermal  energy $k_B T$ at room temperature, $T=300$ K. For sake of comparison,   we also add   the data for the S-shaped LHI (green symbols), indicated by the cyan structure in Fig.~(\ref{modelsPB} A), which is smaller than complex A and complex B.

 As for the GSB complexes, the maximal dipole strength decreases with disorder Fig.~(\ref{PB7} A), even if it is still much larger than one  for $W=k_BT$. On the other hand, the thermal coherence length shows a large  robustness to disorder, Fig.~(\ref{PB7} B) up to $W=k_BT$. 
In Fig.~(\ref{PB7} C,D) the maximal dipole strength (panel C) and the  thermal coherence length (panel D) are shown at $W=0$ and $W=k_BT$  as a function of the number $N$ of chlorophyll molecules contained in the  S-shaped LHI and complexes A and B. 

Both figures of merit  grow as a larger portion of the photosynthetic antenna is considered. Note that in presence of disorder, the maximal dipole strength grows slower with $N$ compared with the case of zero disorder (panel C). On the hand,  the value of the coherence length is the same for the two values of disorder strength considered, showing its  robustness to disorder (panel D). Nevertheless, as $N$ increases, it shows a tendency to saturate.

\section{Conclusions}
\section{Conclusions}
\label{conclusions}
In this manuscript   large scale simulations of the whole antenna complexes of Green sulfur bacteria and Purple bacteria have been performed using a radiative non-Hermitian Hamiltonian, which, at variance with the standard Frenkel dipole Hamiltonian, is valid beyond the small volume limit.  
The largest Green sulfur antenna complex considered was composed of \numprint{132840} Bchl\textit{c} molecules, arranged in three adjacent cylinders, each composed of 4 concentric cylinders of different radii. The size of the largest complex was $148.57 \ \mbox{nm} \times 18.6 \ \mbox{nm} \times 61.8 \  \mbox{nm}$, comparable with the size of natural antenna complexes.  For the Purple bacteria  the whole chromatophore was considered, which is a spherical structure of $60 \ \mbox{nm}$ of diameter and composed of \numprint{4113} Bchl\textit{a} molecules. 

While smaller portions of these complexes have been widely analyzed in literature and superradiance has been found both theoretically and experimentally, an open question was whether  superradiant effects will be  enhanced in the whole complex. Our analysis has shown that the maximal superradiant decay width  in the largest Green sulfur bacteria complex is ten times larger than in a single wall cylinder of the same length (see Fig.~(\ref{f4}) panel A), while in the Purple bacteria chromatophore, it is about six times larger than in the single LHI complex and about ten times larger than the single LHII complex (see Fig.~(\ref{PB7}) panels A and B). 
This proves that the bare structure of the whole antenna complexes is able to support an enhanced  superradiant response. This is not trivial since superradiance critically depends on the arrangement of the emitters and it can be easily quenched. Also the effect of realistic values of  static disorder and thermal noise  have been  investigated. The cooperative effects in large complexes have been found to be robust even with disorder and noise levels comparable with ambient conditions.  In particular, the thermal coherence length in the largest Green sulfur bacteria complex is between two and three times larger than in a single wall cylinder of the same length, see Fig.~(\ref{f4} panel B) while in the Purple bacteria chromatophore the thermal coherence length is approximately two  times larger than in the single (S-shaped) LHI complex.

In analyzing different antenna complexes, we  compared three distinct Hamiltonian models: the Dipole Hamiltonian, applicable in the small volume limit; the Hermitian Hamiltonian, valid when the decay widths are significantly smaller than the mean level spacing; and the full radiative non-Hermitian Hamiltonian model.
For Purple bacteria chromatophores, all models yield consistent results due to the system's small size (e.g., 60 nm) relative to the absorbed wavelength ($\lambda \approx 800 \ \mbox{nm}$) and the limited number of Bchl molecules present. However, for the largest complexes considered in Green sulfur bacteria, only the non-Hermitian Hamiltonian model proved effective in providing reliable results about the eigenstates decay widths. Indeed consistent differences have been found in the largest complex analyzed (GSB complex C). The origin of such differences are not trivial. Indeed they are not due to the fact that the three Hamiltonian models considered (DH, HH, NHH) are characterized by different coupling strengths between the chlorophyll molecules, see \autoref{average-coupling}, but to the fact that the decay widths in the largest system overlap leading to a different distribution of the dipole strengths among the eigenstates. This idea is supported by our analysis of the resonances overlap presented in this manuscript. This shows that the full non-Hermitian Hamiltonian is essential to analyze light matter interaction in molecular aggregates even in the  small volume limit when resonances overlap.

The presence of very bright  (superradiant) states in the whole photosynthetic antenna could shed new light in  understanding the functionality of such complexes. 
In particular, the presence of superradiance and subradiance plays an important role for the efficiency of the energy transfer in these aggregates~\cite{mattioni2021design,kassal2}. Let us also remark that  the system dynamics strongly depends on the choice of the initial state which is mainly  determined by the most superradiant states, see~\autoref{AppD} for a more detailed discussion about the delocalization of superradiant states in PB and GSB antennae.

Our theoretical findings could inspire experimental validation of the extent of cooperative response in natural complexes  predicted in this manuscript. To address these experiments, spectroscopy methods can be employed to characterize the size dependence of  the optical response, comparing different portions of the  photosynthetic antennae  extracted from Green and Purple bacteria, with the whole photosynthetic structures. In particular,  the radiative decay widths obtained through time resolved fluorescence spectroscopy and  the quantum yield obtained  from spectrofluorimetric measurements conducted at different temperatures (from $77 \  \mbox{K}$ to room temperature) could provide a clear signature of cooperative effects present in large photosynthetic aggregates. 

The cooperative properties, akin to those observed in photosynthetic aggregates, are inspiring numerous proposals for engineering artificial  devices for light-harvesting and clean energy production~\cite{scullyPRL,scully1,scully2,creatore,erik2017,superabsorb,zhang, guideslide}. Thus, our study  could significantly impact the development of quantum devices for photon detection and light harvesting.

\begin{acknowledgement}
We acknowledge useful discussion with M. Trotta and Anna Painelli. We also acknowlege  P. Bolpagni  and F. Airoldi for their valuable contribution in the initial stage of this work. We acknowlege L. Pattelli for managing the computer code on one of the LENS clusters and G. Mazzamuto for his help to optimize numerical simulations. E. Pini is kindly acknowledged for help with the simulations. This research was supported in part by grants NSF PHY-1748958 and PHY-2309135 to the Kavli Institute of theoretical Physics (KITP). 
G. L. C. acknowledges financial support from PNRR MUR project PE0000023-NQSTI.
The authors would like to acknowledge the High Performance Computing Center of the University of Strasbourg for supporting this work by providing scientific support and access to computing resources. Part of the computing resources were funded by the Equipex Equip@Meso project (Programme Investissements d'Avenir) and the CPER Alsacalcul/Big Data.
The authors also thank the Institute of Physics BUAP Mexico for allowing the access to their computational resources.
\end{acknowledgement}
\clearpage

\begin{suppinfo}

Additional details and methods, including further  numerical and experimental data are available in the  {\it Supporting Information}.

\end{suppinfo}

\bibliography{bibliography}

\makeatletter\@input{suppinfo.aux.tex}\makeatother

\end{document}

\section{Inverse-Mollweide projection}
\label{AppA}

We use the inverse-Mollweide projection~\cite{snyder1987map} 
to map planar regions to spherical patches. The Mollweide projection from a sphere onto a plane is an area-preserving projection that generates the familiar rendering of the Earth globe where circles of latitude are mapped onto parallel lines on the plane. 
In order to model the PB chromatophore of radius $R$, see Sec.~\ref{PBmodel-vesicle} in the main text,  we use  the following mapping:

\begin{subequations}
 \begin{equation}
 \beta = \arcsin \left( \frac{x}{\sqrt{2}(R+\Delta z)} \right)  
 \end{equation}
 \begin{equation}
\xi = \frac{\pi y}{2 \sqrt{2} (R+ \Delta z) \cos \beta}     
 \end{equation}
 \begin{equation}
     \theta = \arcsin \left( \frac{\sin 2\beta + 2 \beta}{\pi}    \right)
 \end{equation}
 \label{Molleweide1}
\end{subequations}

\begin{equation}
\begin{cases}
\begin{aligned}
x^\prime & =   (R+\Delta z) \cos \theta \cos \xi    \\
y^\prime & =   (R+\Delta z) \cos \theta \sin \xi  \\
z^\prime & =   (R+\Delta z) \sin \theta  
\end{aligned}
\end{cases}
\label{Mollewide2}
\end{equation}
where  $x$ and $y$ are the in-plane coordinates of chlorophyll molecule positions, $\Delta z$ is the vertical shift of the molecules with respect to the plane, $\xi$ and $\theta$ are respectively the longitude and the latitude angles and $x'$, $y'$ and $z'$ are the Cartesian coordinates  on the sphere.

\begin{figure}[!ht]
    \centering
    \includegraphics[scale=0.5]{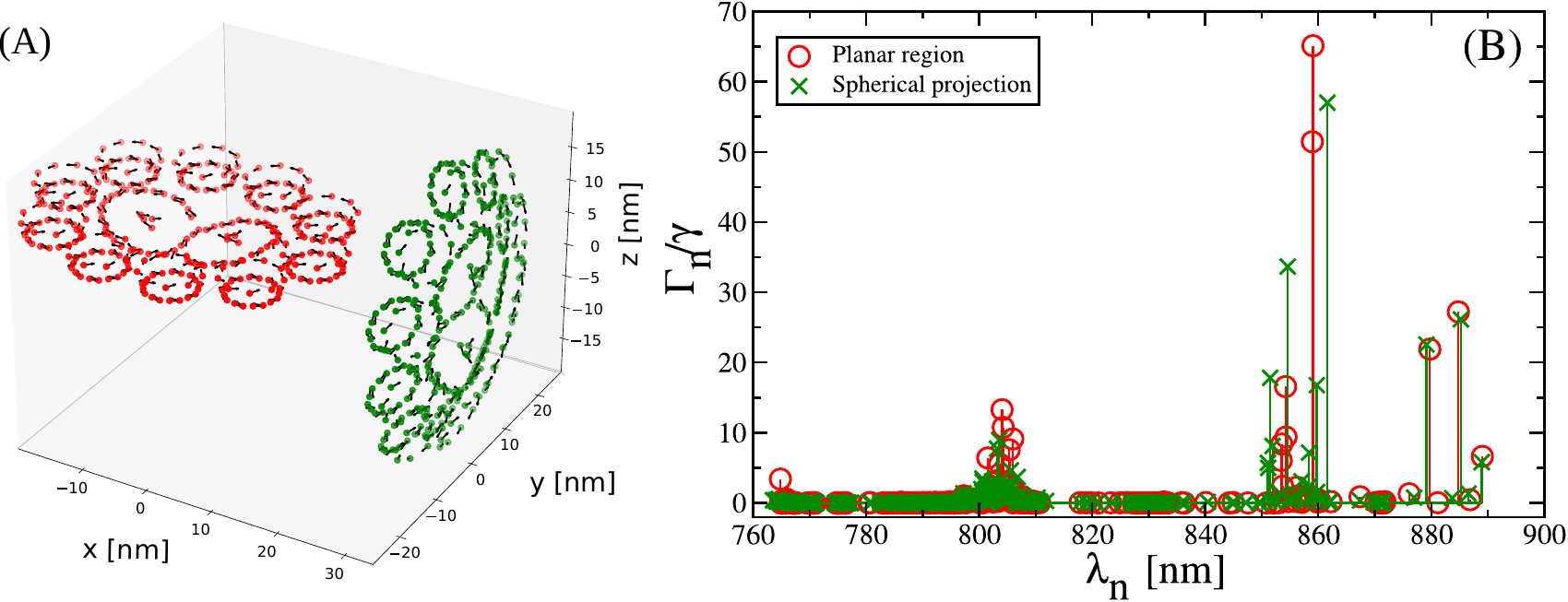}
    \caption{{\it Inverse-Mollweide projection.} Panel (A): Comparison between planar region and spherical area preserving mapping of the PB complex A, see main text, on a spherical cap  with $R=6 \ \mbox{nm}$.  The projection has been   
    obtained by using the inverse Mollweide projection defined in Eq.~(\ref{Molleweide1}-\ref{Mollewide2}).  Panel (B) represents the radiative decay width as a function of the wavelength computed by using the full radiative Hamiltonian in Eq.~(\ref{eq:ham}) for both the systems on planar region (red circles) and on a spherical cap (green crosses). }
    \label{PBprojection}
\end{figure}

Panel A of Fig.~(\ref{PBprojection}) shows the projection of a light-harvesting system comprising 2 RCs and a S-shaped LHI surrounded by 10 LHII complexes B850 and B800 on a spherical cap. 
As expected, the absorption spectrum 
of the projection on the sphere does not differ significantly from that of the planar region, even though small variations that concerns both the positions of the eigenmodes  in the energy spectrum and the decay widths can be observed, see Fig.~(\ref{PBprojection} B).

\newpage
\section{Radiative Hamiltonian parameters}
\label{AppB}
In our model each molecule is represented as a two-level system with an excitation energy $e_0$ and a transition dipole moment (TDM) $\vec{\mu}$. Specifically, we set the excitation energy $e_0$ in $\mbox{ cm}^{-1}$, corresponding to a transition wavelentgth $\lambda_0 = 1/e_0 \times 10^{7} \ \mbox{nm} $, so that the wave vector is $k_0=2 \pi e_0 \times 10^{-8}$~\AA$^{-1}$. Finally if $\vec{\mu}$ is expressed in Debye  and the corresponding  squared transition dipole moment $|\mu|^2$ in   $ \mbox{  \AA}^3 \mbox{  cm}^{-1}$ (for the conversion, see~\cite{Note2}), we can define the radiative decay rate of each molecule $\gamma= 4 |\mu|^2 k_0^3/3 $, corresponding to the radiative lifetime $\tau_\gamma$ (for the conversion, see~\cite{Note3}).

The parameters of the photosynthetic antennae considered here have been taken from literature~\cite{gulli2019macroscopic, prokhorenko2000exciton, prokhorenko2003exciton, kassal2, strumpfer2012excited, mattiotti1}. They are shown in  details respectively in  Tab.~(\ref{tabGSB}) and Tab.~(\ref{tabPB}). 
Moreover the N-N couplings   between dipoles on adjacent sub-units have been shown in Tab.~(\ref{NNcoupling}) for both PB and GSB complexes. 

\begin{table}[!ht]
    \centering
    \begin{tabular}{c c}
    \hline
    \textbf{Site energy}   & $e_0= \numprint{15390} \ \mbox{cm}^{-1}$  \\
    \textbf{Transition wavelength}  & $\lambda_0 \sim 650  \ \text{nm}$\\
    \textbf{Wave vector} &  $ k_0 = 9.670 \cdot 10^{-4 } \ \mbox{\AA}^{-1}$  \\
    \textbf{Dipole moment} &  $ |\vec\mu|=\sqrt{30} \ \mbox{D} $ so that $|\vec\mu|^2 = \numprint{151024} \ \mbox{  \AA}^3 \mbox{  cm}^{-1}$   \\
    \textbf{Decay rate} &    $  \gamma = 1.821 \cdot 10^{-4} \ \mbox{cm}^{-1}   $    \\
    \textbf{Radiative lifetime} &    $ \tau_{\gamma} = 21.15 \ \text{ns} $        \\
    \textbf{N-N couplings} & $\Omega_1 = 618 \ \mbox{cm}^{-1}$ \\
          & $\Omega_2 = 248 \ \mbox{cm}^{-1}$\\
    \hline
    \end{tabular}
    \caption{Parameters for the \textbf{Green sulfur bacteria} photosynthetic antenna. In GSB all the molecules have the same excitation energy $e_0$, while their transition dipole moments have the same magnitude but different orientations in the space. Finally $ \Omega_1$ and $\Omega_2$ are respectively the azimuthal and vertical N-N couplings in GSB complex A.   }
    \label{tabGSB}
\end{table}

\begin{table}[!ht]
    \centering
    \begin{tabular}{c  c  c}
    \hline
        \textbf{Subunit}  & \textbf{Site energy} $[\mbox{cm}^{-1}]$ & \textbf{N-N coupling} $[\mbox{cm}^{-1}]$\\
        \hline
         LHI B875 &  \numprint{12121}  & alt. 300 - 233\\ 
         
         LHII B850 &  alt. \numprint{12458} - \numprint{12654}  & alt. 363 - 320 \\
        
         LHII B800   &  \numprint{12564}  &   /     \\
         
         RC &  P1 \numprint{12180}  &    (P1-P2) 500    \\
            &            &    (P1-B1) -50 \\
            &            &    (P1-B2) -60   \\
            & P2 \numprint{12080}   &   (P2-B2) -50    \\
            &          &       (P2-B1) -60 \\
            & B1 \numprint{12500} &        \\
            & B2 \numprint{12530} &         \\
        \hline
        \textbf{$|\vec{\mu}|$}   &   $10.151 \ \text{D}$ &  \\
        \hline
        \textbf{$\overline{e}_0$} & $ \numprint{12500} \ \text{cm}^{-1}$ & \\
        \hline
    \end{tabular}
    \caption{Parameters for the  \textbf{Purple bacteria}. The table shows the site energies and the nearest-neighbour couplings in each subunits of Purple bacteria light-harvesting complex. The site energies are set to
match the main ﬂuorescence peaks at 800 nm (B800), 850 nm (B850) and
875 nm (LHI). P1, P2 and B1 and B2  represent the four pigments found in each RC. The two strongly-coupled monomers, indicated by P1 and P2, form the special pair, while the other two, B1 and B2, are the accessory pigments in the RC. In the PB antenna, molecules belonging to different subunits have different excitation energies and the nearest-neighbour  (N-N) couplings are given in~\cite{csener2007atomic} and also used in~\cite{mattiotti1} and~\cite{kassal2}. In the LHI and LH2 B850 complexes alternating (alt.) N-N couplings have been used and the excitation energies of the Bchl molecules in the LH2 B850 system are chosen in alternation, as well.    Furthermore the intensity of the transition dipole moment $|\vec{\mu}|$ and the average excitation energy $\overline{e}_0$ computed as the weighted average of the excitation energies of each sub-unit are shown.}
    \label{tabPB}
\end{table}

\begin{table}[h!]
    \centering
    \begin{tabular}{c c c}
       \textbf{Complex}  & \textbf{N-N coupling} $\Omega \ [\mbox{cm}^-1]$& \textbf{Distance} $d \ \mbox[nm]$ \\
       \hline
       
       GSB complex B  & $16$ & $2.1$ \\
       PB complex B & $80$ & $2.2$ \\
       \hline       
    \end{tabular}
    \caption{N-N couplings in PB and GSB complexes. N-N couplings $\Omega$ and distances $d$ between N-N dipoles belonging to adjacent sub-units in GSB and PB complexes are computed by using the HH model and the parameters shown in Tab.~(\ref{tabGSB}) and~(\ref{tabPB}). For GSB complex B  the N-N coupling between dipoles on consecutive cylinders has been considered, while for PB complex B the N-N coupling between dipoles belonging to adjacent LH2 systems is shown. }
    \label{NNcoupling}
\end{table}

\newpage
\section{Cooperativity in GSB antenna}
\label{AppC}

\begin{figure}[!ht]
    \centering
\includegraphics[width=\columnwidth]{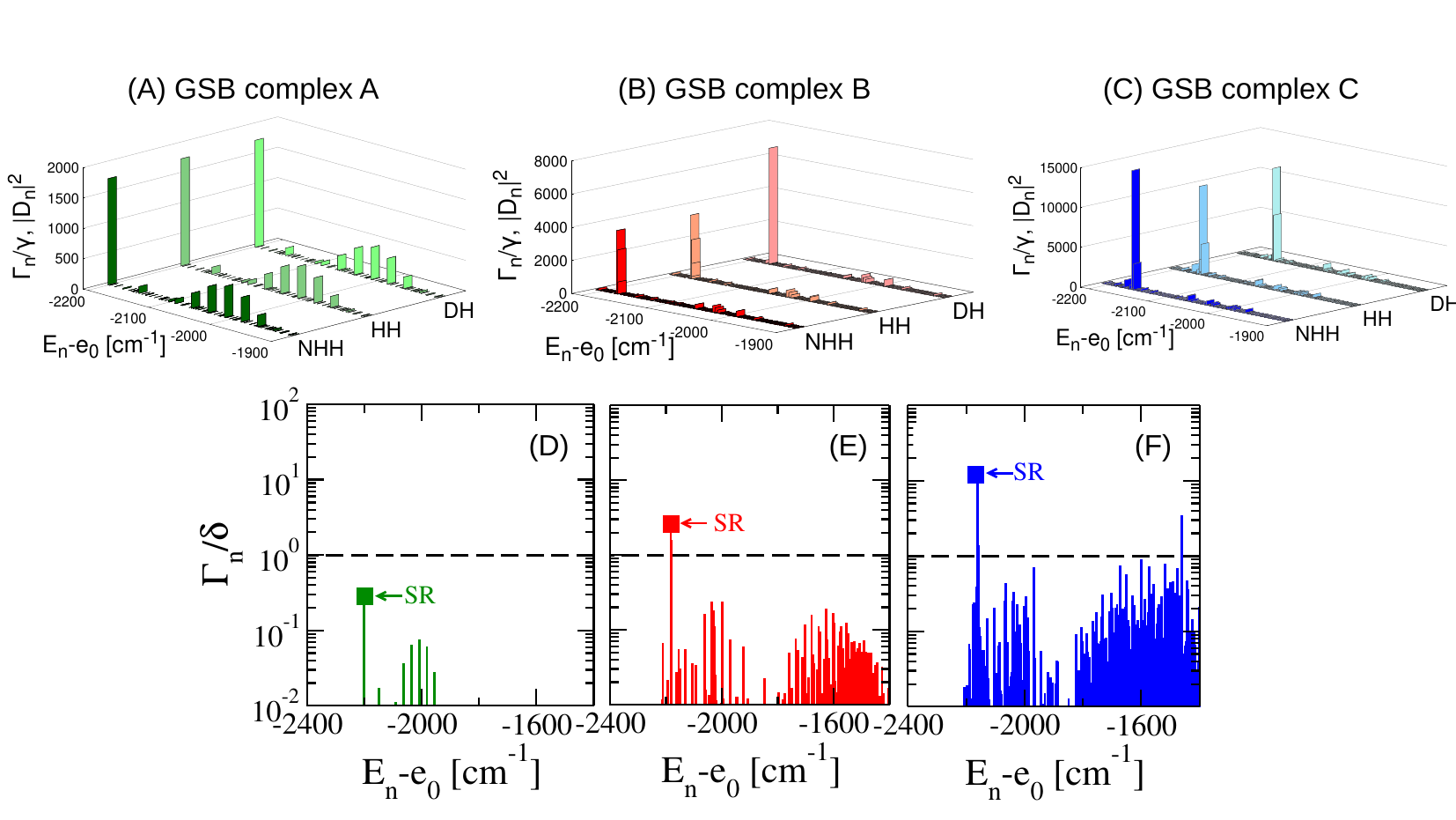}
    \caption{{\it Superradiant states in Green sulfur Bacteria Antennae.}  Complexes A (single cylinder in green colour), B (four concentric cylinders in red colour), C (three adjacent concentric cylinders in blue colour) are shown in the respective panels (Panels A, B, C). Three different Hamiltonian models are compared for each model using different shades
of colour (NHH: non-Hermitian Hamiltonian Eq.~(\ref{eq:ham}, in dark colours), HH: Hermitian Hamiltonian Eq.~(\ref{eq:hreal}, in medium shade of colours), DH: Dipole Hamiltonian Eq.~(\ref{dip}, in light colours). The squared dipole strength $|\mbox{D}_n|^2$, see Eq.~(\ref{eq:dipst}), is shown for the HH and DH models, while the radiative decay width $\Gamma_n/\gamma$ is shown for the NHH model  as a function of the energy $E_n-e_0$. Panels (A-C) show only the lowest part of the energy spectrum where the most superradiant states are located. Results are computed by using a fixed length $L= 82.17 $ nm for each aggregate, which corresponds to 100 rings for each  cylinder. (D-F) These panels show the ratio between the decay width $\Gamma_n$ obtained by diagonalizing the full radiative Hamiltonian (NHH model) in Eq.~(\ref{eq:ham}) and the mean level spacing $\delta$ as a function of the energy $E_n-e_0$ for complexes A (in dark green colour), B (in dark red colour), C (in dark blue colour) with a fixed length $L= 82.17 $ nm. The mean level spacing $\delta$ is computed as the ratio between the energy spectral width and the total number of eigenmodes. Green, red and blue squares indicate the positions in the energy spectra of the most superradiant state for each complex. The horizontal dashed line
    represents the value of the ratio ($\Gamma_n/\delta=1$) above which resonances overlap.}
    \label{GSB4}
\end{figure}

In this section we consider GSB complexes A, B and C, see Sec.~\ref{numericalGSB} in the main text, with a length $L=82.17 \ \text{nm}$, corresponding to $100$ rings for each cylindrical structure. Panels A, B and C of Fig.~(\ref{GSB4}) show the energy spectra computed respectively for a single cylinder (A, in green colour), four concentric cylinders (B, in red colour) and three adjacent cylinders with four concentric rolls each (C, in blue colour). Here the three Hamiltonian models are compared: NHH (in dark colours), HH (in medium shade of colours) and DH (in light colours). For complex A all the models we considered (NHH, HH and DH) give a good description of superradiance, for complex B the DH model is no longer valid, while HH and NHH model are in agreement even if small deviations can be observed, finally for complex C the NHH model differs from the HH model only in the most superradiant states, with  a $20\%$ difference, much smaller than what happens for larger system sizes. This should be compared with Fig.~(\ref{f1}), where the larger system size is shown and where the differences between the HH and NHH model are much larger (close to $60\%$ for complex C) and involve a much larger number of states. For this reason to study the robustness of cooperative effects to disorder and noise for this system size (100 rings) we used the HH model, see discussion in Sec.~\ref{numericalGSB}.  

In panel D of Fig.~(\ref{GSB4}), it is shown that for complexes A and B the non-Hermitian part of the Hamiltonian can be considered perturbatively, while for complexes C only for the most superradiant states the resonances overlap. Again this should be compared with panel (D) of Fig.~(\ref{f1}) where for larger system sizes it is shown that the decay widths of a  much larger number of states overlap. 

\clearpage

\section{Probability distribution of the most superradiant states}
\label{AppD}
Knowing the cooperativity of the antenna complexes can be useful for understanding the initial state to study the energy transfer within these aggregates. Indeed under light illumination, superradiant states are the most excited ones and so they determine the correct initial conditions for the energy transfer.  
To address this issue, here we consider the probability distribution over the sites of the   superradiant states both in the GSB and in the PB antenna.

\subsection{Most superradiant state in the GSB antenna}

\begin{figure}[!ht] 
\centering
\includegraphics[width=\columnwidth]{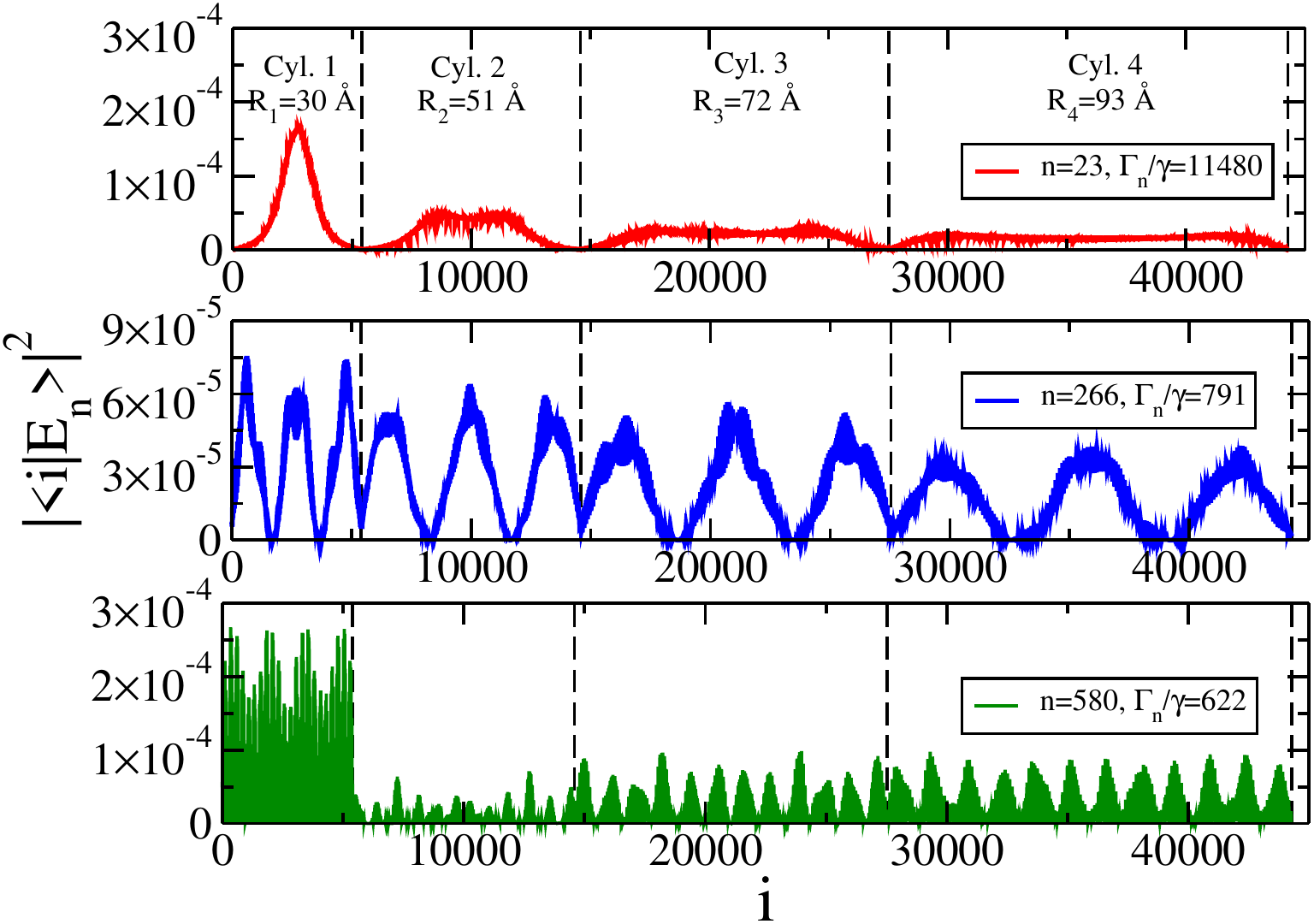}
\caption{{\it Delocalization of the superradiant states in complex B of GSB.} The probability $|\braket{i|E_n}|^2$ of the three most superradiant eigenmodes to be on the $i^{th}$ site of the aggregate is shown. The most superradiant states are respectively the $23^{rd}$ excited state (red continuous line), the $266^{th}$ exc. state (blue continuous line) and the $580^{th}$ exc. state (green continuous line). Here we consider an aggregate with four concentric cylinders (GSB complex B) made of $180$ rings with a length $L=148.57 \ \mbox{nm}$. The vertical black dashed lines define the end of an inner cylinder and the beginning of an outer one. The eigenstates  of the system $\braket{i|E_n}$ are computed by using the radiative  Hamiltonian in Eq.~(\ref{eq:ham}). 
} \label{GSB9}
\end{figure}

 Here we consider GSB complex B with four concentric cylinders, comprising $180$ rings. Fig.~(\ref{GSB9}) shows how the three most superradiant eigenmodes are delocalized on the four concentric cylinders.

\clearpage

\subsection{Most superradiant state in the PB antenna}

\begin{figure}[!ht]
    \centering
    \includegraphics[scale=0.5]{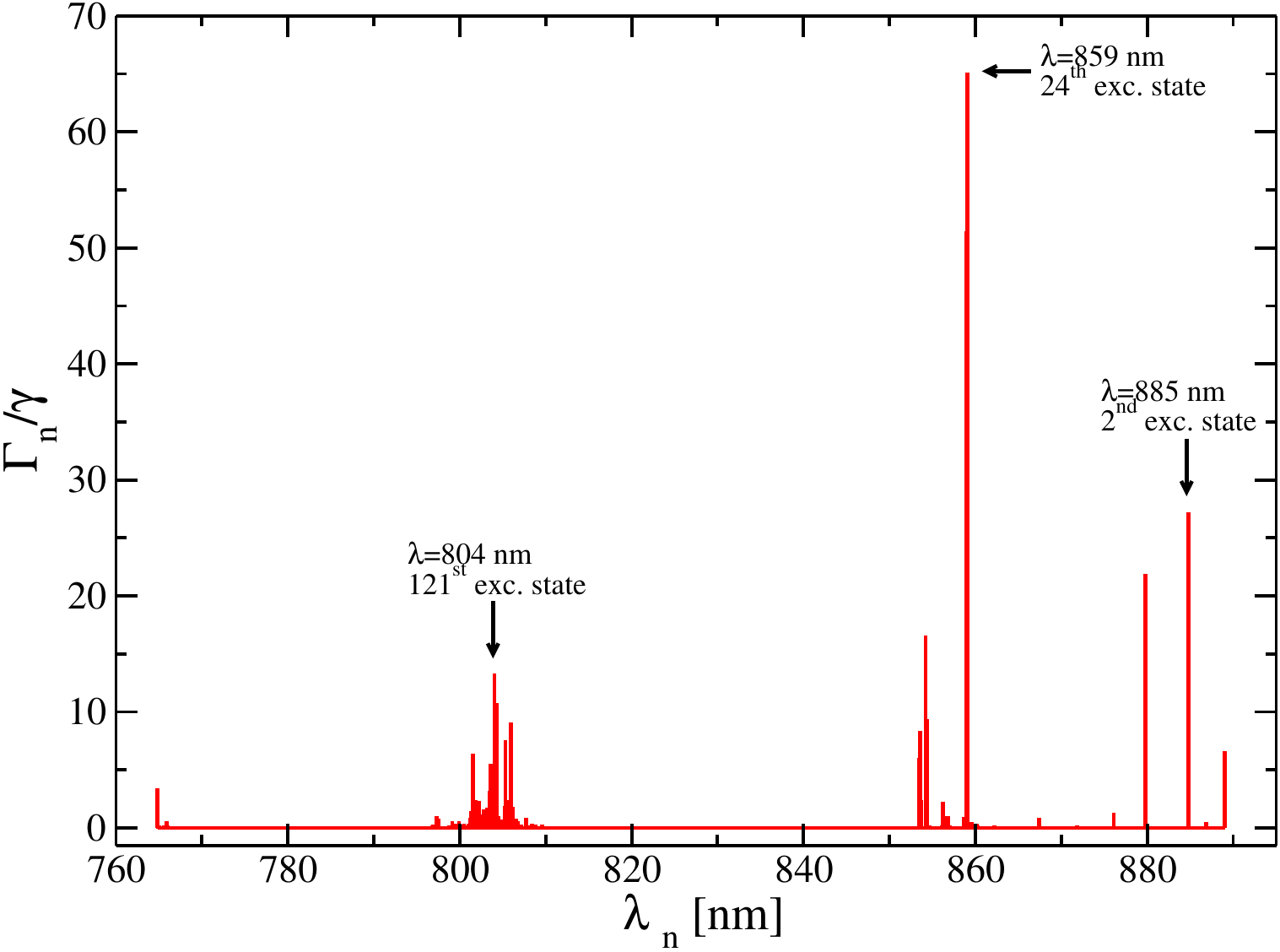}
    \caption{{\it Cooperativity and superradiance in PB light-harvesting systems (complex A).}  Energy spectrum of the light-harvesting system found in Purple bacteria with $N=334$ Bchls formed by a S-shaped LHI complex (B875) with 2 RCs surrounded by 10 LHII rings (B850 and B800) (see panel (A) of Fig.~(\ref{modelsPB})). Radiative decay width  calculated with the NHH model as a function of the wavelength. The square dipole strengths computed by using DH and HH models are not shown since they give similar results.
    Black arrows indicate the positions in the energy spectrum of the most superradiant eigenstates which are delocalized on the subunits of the light-harvesting complex here considered, see Fig.~(\ref{PB2}).
    }
    \label{PB1}
\end{figure}

\begin{figure}[!ht]
    \centering
    \includegraphics[scale=0.5]{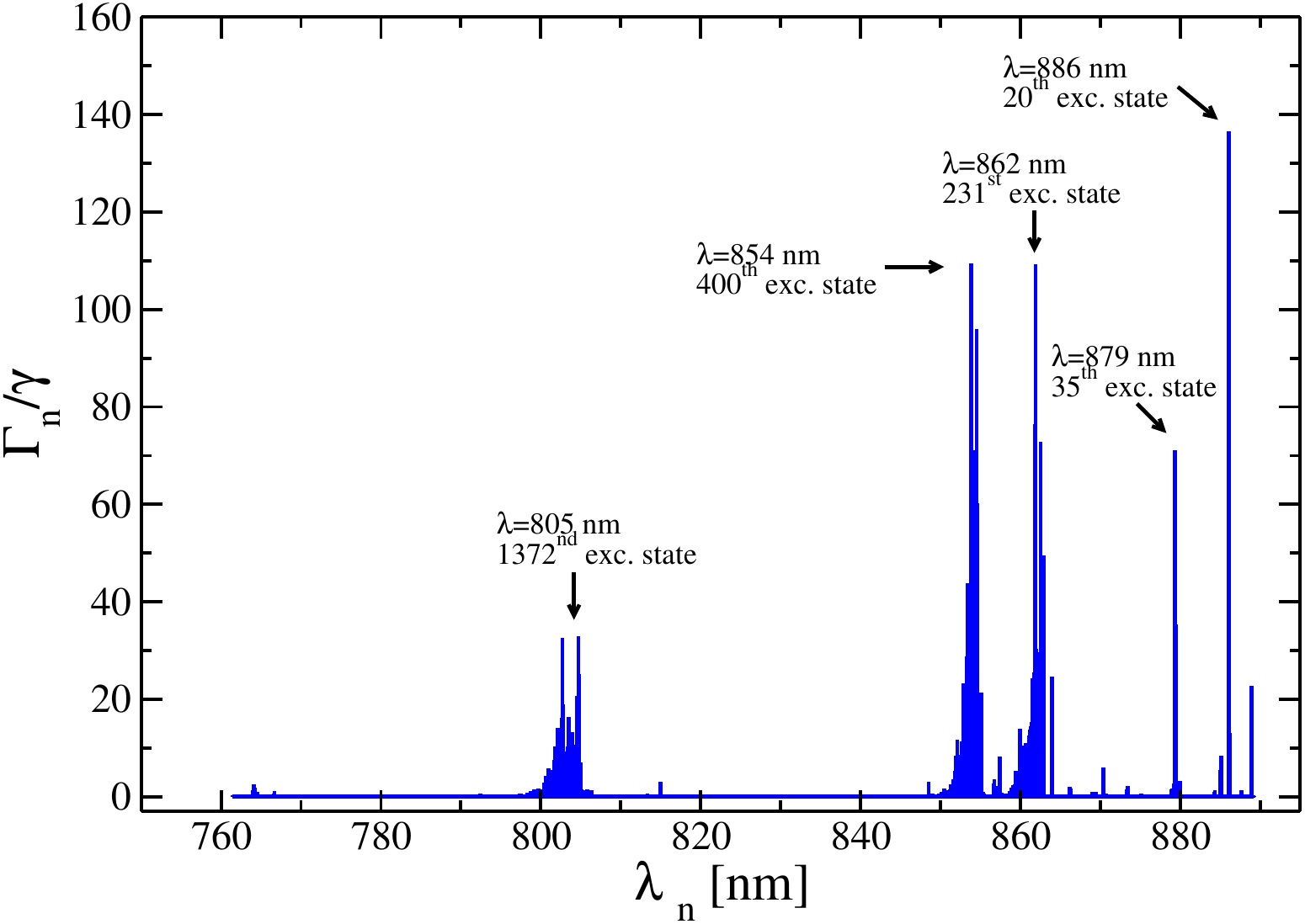}
    \caption{{\it Cooperativity and superradiance in PB chromatophore (complex B).} 
    Energy
spectrum of the chromatophore of Purple bacteria with $N=\numprint{4113}$ Bchls (see panel (B) of Fig.~(\ref{modelsPB})). Radiative decay width calculated with the NHH model as a function of the wavelength. The square dipole strengths computed by using DH and HH models are not shown since they give similar results. Black arrows
indicates the positions in the energy spectrum of the most superradiant eigenstates which are delocalized on the subunits of
the light-harvesting complex here considered, see Fig.~(\ref{PB4}). }
    \label{PBapp}
\end{figure}

Fig.~(\ref{PB1}) and Fig.~(\ref{PBapp}) represent respectively the energy spectra of complexes A (red colour) and B (blue colour) of PB antennae, showing the comparison between the three models we used: NHH (square symbols), HH (circle symbols) and DH (cross symbols). 
The decay width, computed using the NHH model, and the squared dipole strength, computed using both the HH and DH models, are shown as functions of the wavelength in these figures.  
The most superradiant states at approximately $800$, $860$, and $880$ nm are indicated by black arrows.

\begin{figure}[!ht]
    \centering
    \includegraphics[width=\columnwidth]{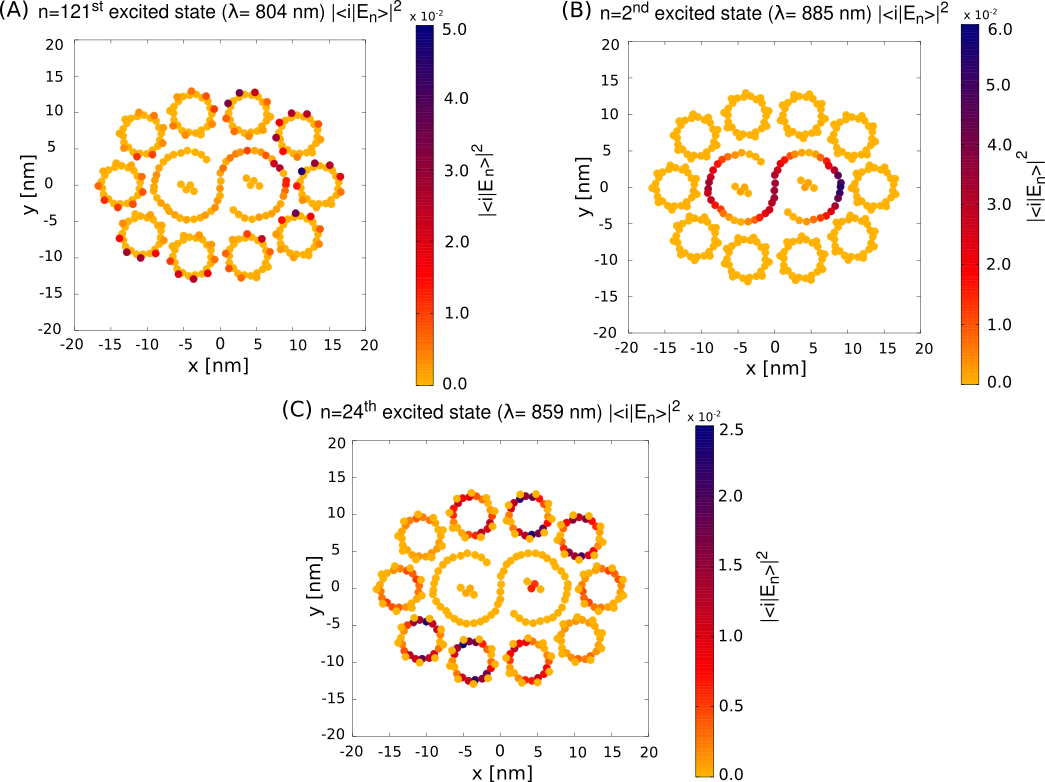}
    \caption{{\it Delocalization of superradiant eigenmodes in complex A of PB.} (Panels A-C) delocalization of the most superradiant eigenstates on the sites of the system. $|\braket{i|E_n}|^2$ is the probability of the $n^{th}$ eigenstate to be delocalized on the $i^{th}$ site. The eigenstates of the system are computed by diagonalizing only the Hermitian part of the full radiative Hamiltonian described in Eq.~(\ref{eq:ham}). See Fig.~(\ref{PB1}) to read the positions in the energy spectrum of the most superradiant states. The colorbar on the right represents the values of the probability of the $n^{th}$ eigenmode to be projected on a site.}
    \label{PB2}
\end{figure}

\begin{figure}[!ht]
    \centering
    \includegraphics[scale=0.65]{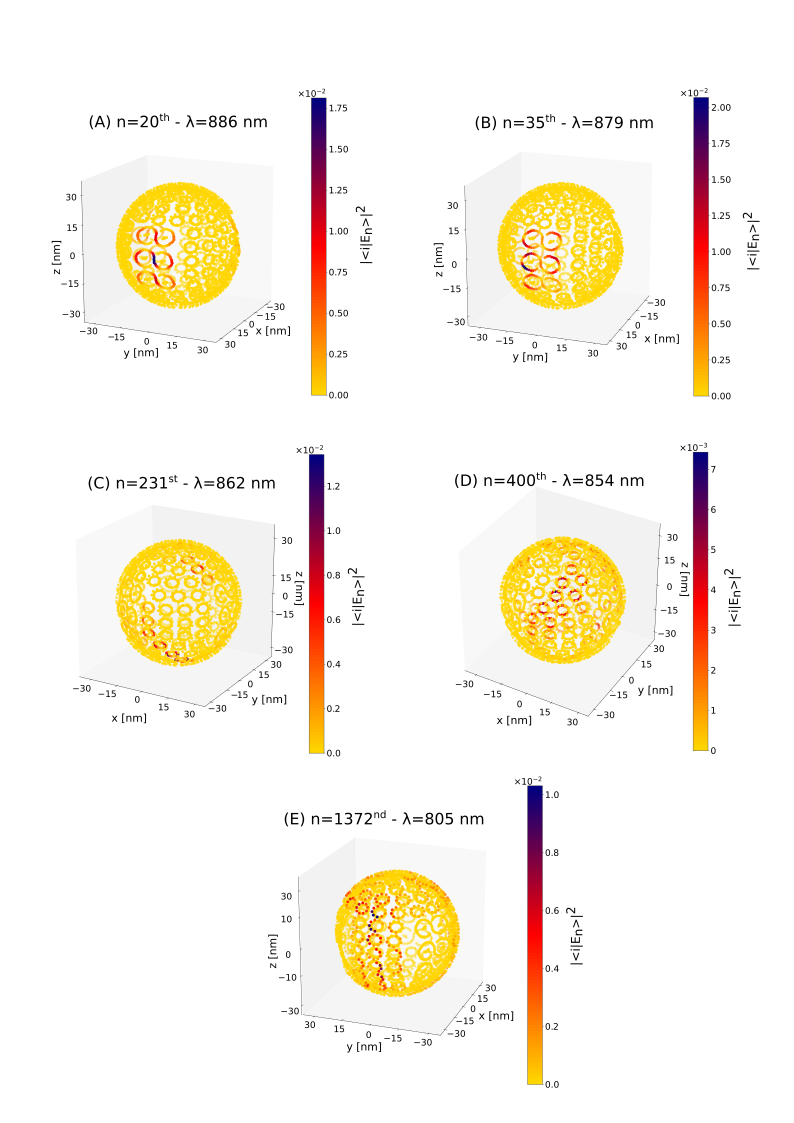}
    \caption{{\it Delocalization of superradiant eigenmodes in complex B of PB.} (Panels A-E) probability of the superradiant eigenstates $|\braket{i|E_n}|^2$ to be projected on the sites.  The eigenstates of the system are computed by diagonalizing only the Hermitian part of the full radiative Hamiltonian described in Eq.~(\ref{eq:ham}). The colorbar on the right represents the values of the probability of the $n^{th}$ eigenmode to
be projected on a site. See Fig.~(\ref{PBapp}) to read the positions in the energy spectrum of the
most superradiant states. }
    \label{PB4}
\end{figure}

Fig.~(\ref{PB2}) and Fig.~(\ref{PB4}) show the probability of the most superradiant eigenstates being delocalized across the sites of the system. Fig.~(\ref{PB2}) refers to PB complex A, while Fig.~(\ref{PB4}) to PB complex B. Each panel in both figures represents the projection  of one of the most superradiant states, identified by black arrows in their respective energy spectra. These figures reveal that in both complexes, superradiant states at $800$ nm are primarily delocalized over the sites of the LHII B800 rings, those at $860$ nm are delocalized over the LHII B850 rings, and finally, superradiant states at around $880$ nm are delocalized over the LHI system.

\clearpage

\section{PB chromatophore structure}
\label{AppE}
\begin{figure}[!ht]
    \centering
    \includegraphics[width=\columnwidth]{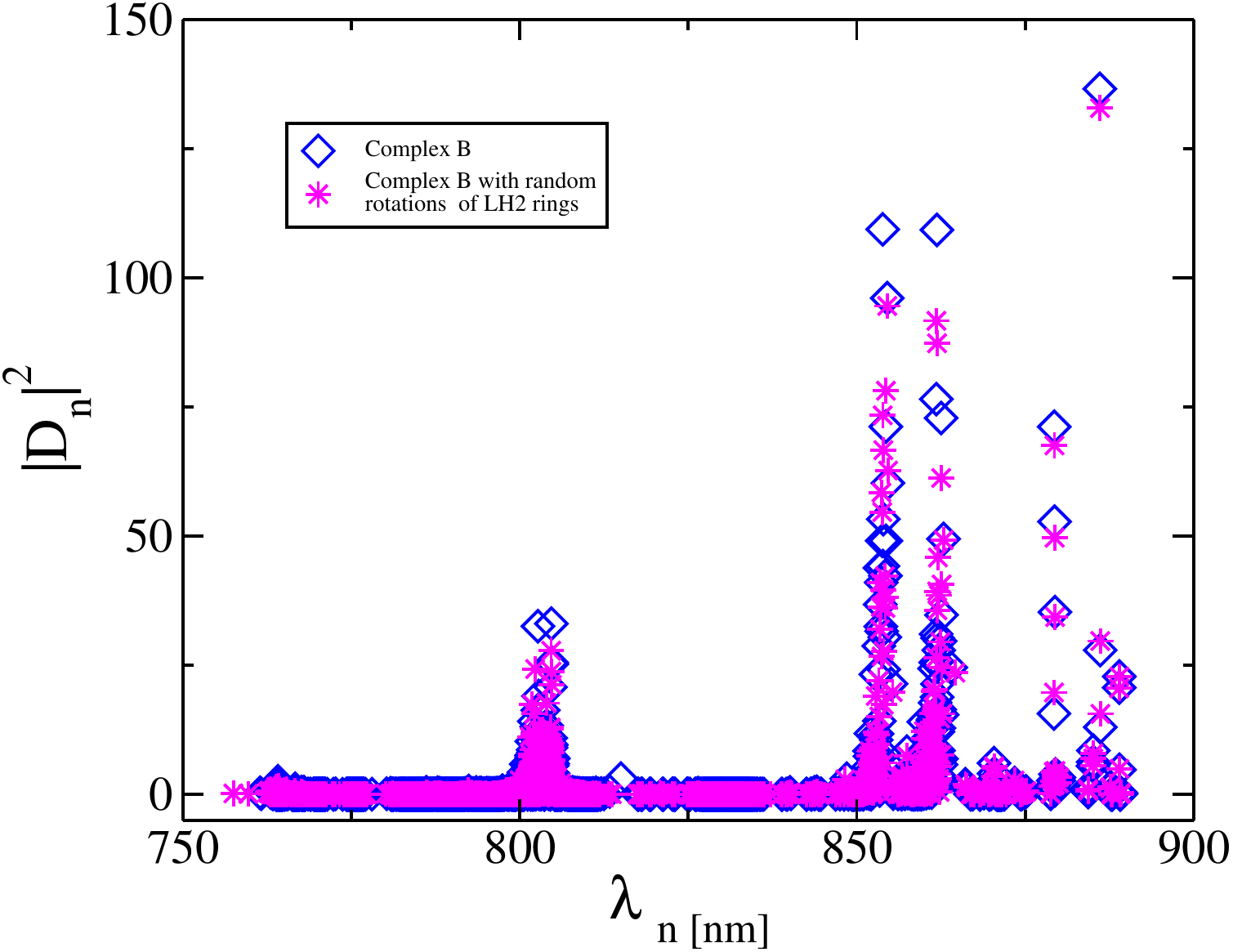}
    \caption{{\it Random rotations of LHII light-harvesting systems in the chromatophore  of PB.} Comparison between the energy spectra computed for the chromatohore shown in the text (blue diamond) and the chromatophore  with LHII rings  rotated around the axis joining the origin of the vesicle and the center of the ring  by a random angle (magenta stars). The dipole strength  as a function of the wavelength have been computed with the DH model shown in Eq.~(\ref{dip}). }
    \label{PBrotationLHII}
\end{figure}

Here we consider a different model more for PB chromatophore. A vesicle (PB complex B) where all the LHII rings are randomly rotated around the axis joining the center of the chromatophore with the center of each LHII ring is studied. Fig.~(\ref{PBrotationLHII}) shows the comparison between the energy spectra computed for a chromatophore with randomly rotated LHII rings (magenta stars) and without rotations (blue diamonds) as a function of the wavelength. Results are obtained by using the DH model (which is valid for the PB chromatophore). It can be observed that the differences between the two cases are not very large. For this reason we did not implement random rotations of the LHII complex in the PB chromatophore model considered in the main text. 

\clearpage
\section{Robustness of Superradiance with respect to random dipole orientation}
\label{AppF-GSB}
The main purpose of this manuscript was to study cooperative effects at large scale, for realistic photosynthetic antennae sizes. Our findings reveal that as the portion of the antenna analyzed is increased, also the cooperative effects are enhanced. To show that this is not a trivial effect, we here consider how cooperativity is affected when the dipole orientation are partially randomized. 
For this reason we compare the results shown in Sec.~\ref{numerical} for the PB and GSB complexes with these new models.
For the GSB antenna, a system with four concentric cylinders with $100$ rings is considered. All the dipoles have the same positions as the complex B presented in Sec.~\ref{subsec:GSBmodel}. The dipole  orientation is randomized in all cylinders but one (the one with radius of 7.2 nm which is the closest in diameter to GSB complex A).  On the other hand, for the  PB antenna we consider a system where all  dipoles have   the same positions as in the chromatophore presented in Sec.~\ref{PBmodel-vesicle}, nevertheless the dipole orientation  are randomized for all molecules except the molecules belonging to a single LHI $+$ 2RCs complex (which have been chosen at random among the 9 LHI+RC complexes present on the chromatophore). 

\begin{figure}[!ht]
    \centering
    \includegraphics[width=\columnwidth]{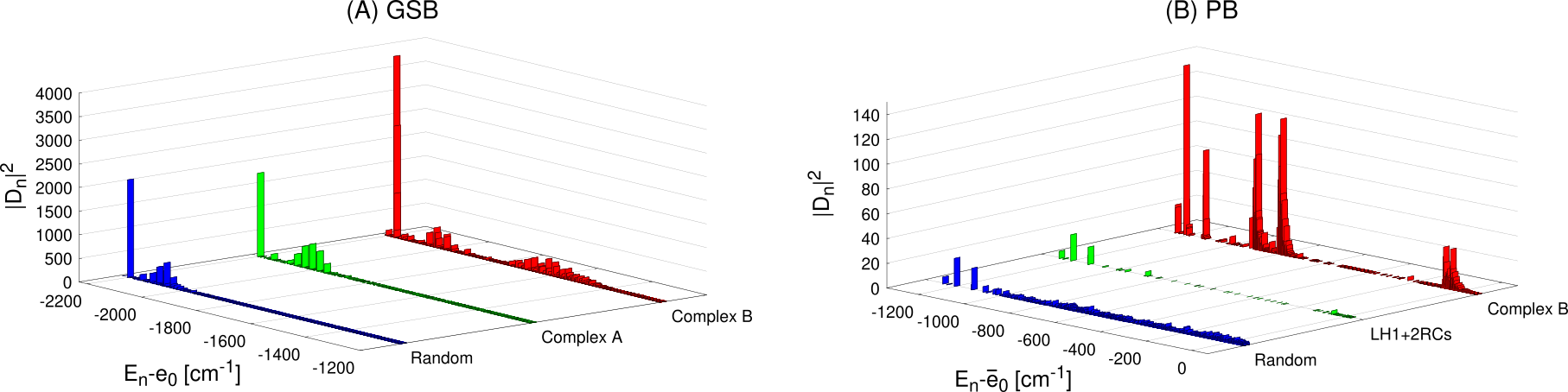}
    \caption{{\it GSB and PB spectra: the effect of randomness on superradiance.    } The squared dipole strength as a function of the energy computed with the HH model is represented in both panels A and B. (A) GSB complex with a fixed length $L=82.17 \ \mbox{nm}$, corresponding to $100$ rings. GSB complex B (in red colour) comprises four concentric cylinders, while GSB complex A (in green colour) is made of a single wall. In both of them the dipoles orientations are defined in the main text, see Sec.~\ref{subsec:GSBmodel}. GSB random dipoles (in blue colour) have the same structure as in the GSB complex B, but with random orientation except for the  the $3^{rd}$ cylinder with radius $R=7.2 \ \mbox{nm}$. (B) PB complex spectra are shown. PB complex B (in red colour) represents the chromatophore described in Sec.~\ref{PBmodel-vesicle}, while PB LHI $+$ 2RCs (in green colour) comprises a single S-shaped LHI with two RCs projected on a spherical cap. Finally, PB random dipoles (in blue colour) represent the chromatophore, where all dipoles have random orientations, except  those  belonging to a single LHI $+$ 2RCs.  }
    \label{random}
\end{figure}

Fig.~(\ref{random}) shows the dipole strength as a function of the energy computed with the HH model  for both PB and GSB complexes. For both models the comparison between the randomized system and the realistic complexes (GSB complex B and PB complex B) shows that  the presence of randomized dipoles strongly decreases  superradiance. Nevertheless,  we can observe that superradiance in random complexes is not completely suppressed. Indeed,   superradiant states survive and are characterized by dipole strengths  comparable with those  found in the portion of the antenna whose dipoles have not been randomized.  Probably this is due to the fact that the coupling between different sub-units in the GSB and PB antennae are not very strong. Indeed, in general surrounding an ordered structure, which presents superradiance, with dipoles having random orientation could easily quench superradiance. 

\clearpage

\section{Absorption spectra in GSB and PB aggregates}
\label{Spectra}
\begin{figure}[h!]
    \centering
    \includegraphics[width=\columnwidth]{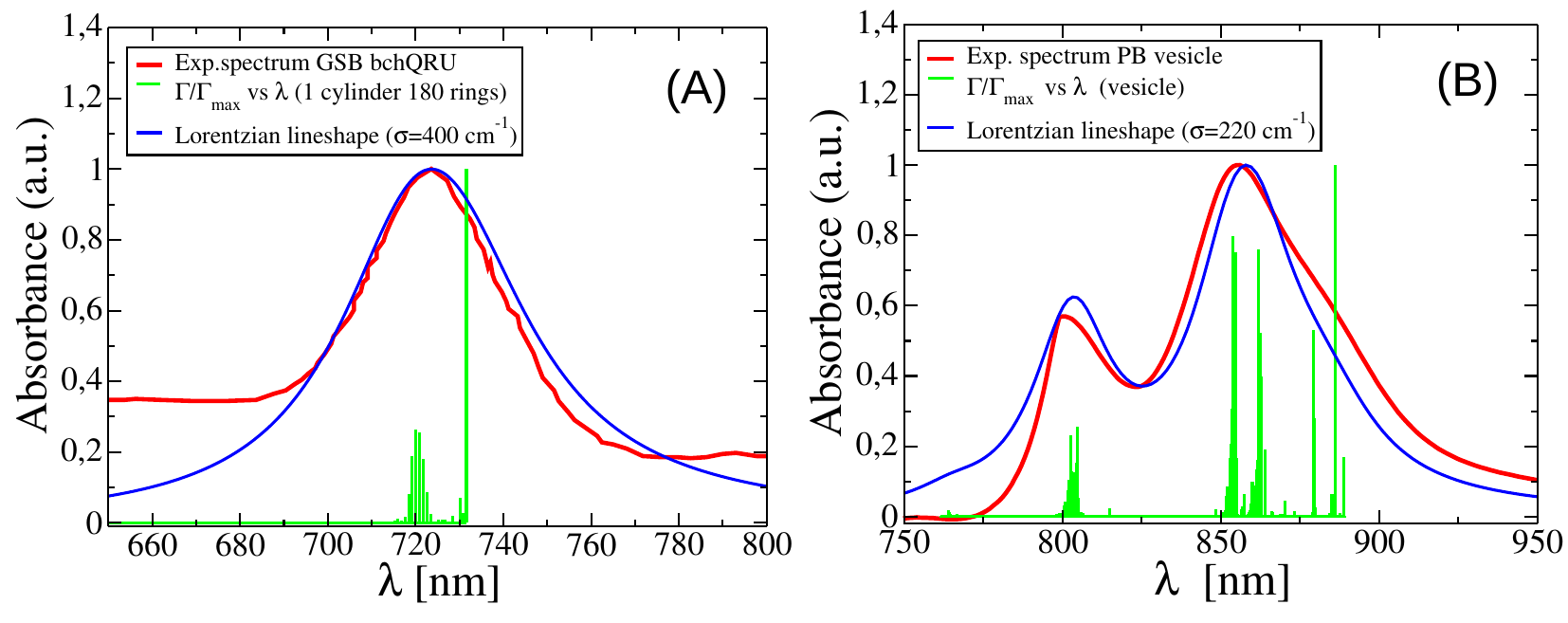}
    \caption{\textit{Absorption spectra of GSB and PB complexes.} Experimental and numerical normalized absorption spectra for GSB bchQRU triple mutant and PB \textit{Rhodobacter sphaeroides} are compared. Panel (A): the experimental absorption spectrum of whole cells of GSB bchQRU mutant (red solid line) taken from Ref.~\cite{chew2007bacteriochlorophyllide}  is compared with the numerical spectrum for a single cylinder with 180 rings (blue solid line) obtained with Eq.~\ref{abs-marcus1}, assuming a Lorentzian lineshape with homogeneous broadening $\sigma=400 \ \mbox{cm}^{-1}$. The numerical spectrum has been shifted  by $27$ nm to the blue to match the position of the experimental absorption maximum.  Panel (B): experimental data  (red solid line) provided by Ref.~\cite{Note4} % Prof. M. Trotta group (CNR Bari) 
    are compared with the numerical spectrum for a chromatophore (blue solid line) obtained with Eq.~\ref{abs-marcus1}, assuming a Lorentzian lineshape with homogeneous broadening $\sigma=220 \ \mbox{cm}^{-1}$. 
    In both panels (A-B) the light green solid line represents the radiative decay rate as a function of the wavelength computed with the NHH model.}
    \label{abs}
\end{figure}

In this section a comparison between our models and the experimental results in both PB and GSB aggregates is considered. In  Fig.~(\ref{abs}) we show that our theoretical model reproduces the basic features of the experimental absorption  spectra.\\

The absorption spectrum indicates the amount of incident electromagnetic radiation absorbed
by the Bchl chromophores in the aggregate, between a range of energies or frequencies~\cite{babcock2024ultraviolet}. Following Ref.~\cite{renger2002relation}, the linear absorption spectrum can be defined as the real part of the Fourier–Laplace transform of the dipole–dipole correlation function, also known as the lineshape function, which can be expressed as a function of the energy:
\begin{equation}
A(E) = K \sum_{j} \Gamma_j D_j(E)\,
    \label{abs-marcus1}
\end{equation}
where $\Gamma_j$ is the decay corresponding to each $j^{\mbox{th}}$ eigenstate of the Hamiltonian, K is a normalization factor and $D_j(E)$ is the lineshape function. Here we assume only homogeneous broadening, so a Lorentzian lineshape function centred at $E_j$ is considered:
\begin{equation}
   D_j(E) = \dfrac{\sigma}{(E-E_j)^2+\sigma^2}\,
    \label{abs-marcus2}
\end{equation}
where the parameter $\sigma$ is the strength of the homogeneous broadening.

\clearpage

\section{Analysis of the three Hamiltonian models}
\label{average-coupling}

In this section we analyze  in more detail the difference between the three Hamiltonian models considered to study the photosynthetic antennae.  In particular, the average coupling in all the complexes (GSB complexes A,B,C and PB complexes A,B) analyzed in the manuscript for the  three different Hamiltonian models is studied. The average coupling $\braket{J}$ has been computed in the following way: 
\begin{equation}
 \braket{J} = \dfrac{1}{N} \sum_{i=1}^N \sqrt{ \sum_{j \ne i}  |H_{i,j}|^2}, 
    \label{eq:coupling}
\end{equation}

where N is the system size, while $H_{i,j}$ is the Hamiltonian matrix element $H_{i,j}=\bra{i}H\ket{j}$.

\begin{table}[h!]
    \centering
    \begin{tabular}{|c|c|c|c|c|}
    \cline{3-5}
       \multicolumn{2}{c|}{}   & NHH & HH & DH \\
         \multicolumn{2}{c|}{}     &$[\mbox{cm}^{-1}]$ & $[\mbox{cm}^{-1}]$ & $[\mbox{cm}^{-1}]$ \\
         \hline
     GSB & complex A & 981.08233 & 981.08233 &981.04971 \\
         \cline{2-5}
       & complex B & 982.53042  & 982.53042& 981.04971  \\
         \cline{2-5}
         & complex C & 982.53956  & 982.53956 &  982.49831 \\
         \hline
         \hline
        PB & complex A & 394.31019 &   394.30997 & 394.30079  \\
         \cline{2-5}
        & complex B & 393.40191  & 393.40191 & 393.39299  \\
         \hline
    \end{tabular}
    \caption{Average couplings $\braket{J}$ (see Eq.~\ref{eq:coupling}) obtained with the three Hamiltonian models (NHH, HH and DH) for all the complexes we consider in the manuscript (GSB complexes A, B, C with length L=150 nm and PB complexes A, B). }
    \label{tab-coupling}
\end{table}

\begin{table}[h!]
    \centering
    \begin{tabular}{|c|c|c|c|c|c|c|c|}
    \cline{3-8}
    \multicolumn{2}{c|}{} & \multicolumn{3}{c|}{$E_{max}-E_{min}$}  & \multicolumn{3}{c|}{$E_{min}-e_0$}\\
    \multicolumn{2}{c|}{} & \multicolumn{3}{c|}{$[\mbox{cm}^{-1}]$}  & \multicolumn{3}{c|}{$[\mbox{cm}^{-1}]$}\\
    \cline{3-8}
        \multicolumn{2}{c|}{}  & NHH & HH & DH  & NHH & HH  & DH \\
    \hline
     GSB   & complex A & 3663.92 & 3663.92 & 3661.67  & -2207.69  &  -2207.69 & -2205.47 \\
    \cline{2-8}
        & complex B & 3681.73 & 3668.2 & 3667.95  &  -2216.46 &  -2210.87 & -2210.64 \\
    \cline{2-8}
        & complex C & 3668.41 & 3668.41 & 3667.99  & -2211.08  &  -2211.08 & -2210.68 \\
        \hline
        \hline
     PB   & complex A & 1826.67 & 1826.67 & 1826.61  &  -1252.79 &  -1252.79 & -1252.71 \\
    \cline{2-8}
          & complex B & 1886.7 & 1886.7 &  1886.7 & -1254.88  & -1254.88  & -1254.88 \\
    \hline
    \end{tabular}
    \caption{Spectral width $E_{max}-E_{min}$ and red-shifted GS energy $E_{min}-e_0$ computed with the NHH, HH and DH models for
each complex here considered (GSB complexes A, B, C with length L=150 nm and
PB complexes A, B). $e_0$ represents the excitation energy of the single
molecule.}
    \label{tab-spectral}
\end{table}

Tab.~(\ref{tab-coupling}) shows that the coupling are almost identical. Also the spectral widths and the lowest excitonic energy obtained with the three different Hamiltonian models for all the complexes considered are almost identical, see Tab.~(\ref{tab-spectral}).

The reason for these results is explained here below. 
\begin{itemize}
    \item NHH and HH differ only by the non-Hermitian part in the Hamiltonian. Its contribution is very small, of the order of $10^{-4} \ \mbox{cm}^{-1}$, see~\cite{gulli2019macroscopic}, while the Hermitian part gives a maximal contribution of about $600  \ \mbox{cm}^{-1}$.
    \item The DH and the other models have similar couplings since the maximal system size considered, which is about 150 nm, is smaller than the transition wavelength $\lambda_0$, which is 650 nm. Thus, so as far as the couplings are concerned, the small volume limit is a good approximation.
\end{itemize}

Nevertheless the differences we have found between the three Hamiltonian models regard the lifetime of the eigenstates (decay widths) and not their spectral width. These are not simply due to the coupling strengths but they depend on how the dipole strength is redistributed among the eigenstates. This redistribution is very sensitive to the overlap of the eigenstates decay widths (resonance overlap criterion). This is explained in the text, see Fig.~(\ref{f1}) (panels D,E,F), Fig.~(\ref{PB3}) (panels: C,D) and Fig.~(\ref{GSB4}) (panels: D,E,F).

\clearpage

\bibliography{bibliography}

\makeatletter\@input{main.aux.tex}\makeatother